\documentclass[twocolumn]{aastex63}

\usepackage{color}
\usepackage{amsmath}
\usepackage{booktabs}
\usepackage{enumerate}
\usepackage[version=3]{mhchem}
\bibpunct{(}{)}{;}{a}{}{,}
\definecolor{revised}{rgb}{0,0.2,0}
\definecolor{ToBeRevised}{RGB}{255,147,0}
\definecolor{ToBeRevisedLater}{rgb}{0,0,0}
\definecolor{PleaseRevise}{rgb}{1,0,0}
\definecolor{others}{RGB}{255,147,0}

\shorttitle{$K$-band High-Resolution Spectroscopy of Embedded Massive Protostars}
\shortauthors{Hsieh et al.}

\begin{document}

\title{$K$-band High-Resolution Spectroscopy of Embedded High-Mass Protostars}

\correspondingauthor{Tien-Hao Hsieh}
\email{thhsieh@mpe.mpg.de }

\author[0000-0002-5507-5697]{Tien-Hao Hsieh}
\affiliation{Institute of Astronomy and Astrophysics, Academia Sinica
11F of Astronomy-Mathematics Building
No.1, Sec. 4, Roosevelt Rd, Taipei 10617, Taiwan}
\affiliation{Max-Planck-Institut f\"{u}r extraterrestrische Physik, Giessenbachstrasse 1, D-85748 Garching, Germany}

\author{Michihiro Takami}
\affiliation{Institute of Astronomy and Astrophysics, Academia Sinica
11F of Astronomy-Mathematics Building
No.1, Sec. 4, Roosevelt Rd, Taipei 10617, Taiwan}

\author{Michael S. Connelley}
\affiliation{Institute for Astronomy, University of Hawaii, 2680 Woodlawn Drive, Honolulu, HI 96822, USA}

\author{Sheng-Yuan Liu}
\affiliation{Institute of Astronomy and Astrophysics, Academia Sinica
11F of Astronomy-Mathematics Building
No.1, Sec. 4, Roosevelt Rd, Taipei 10617, Taiwan}

\author{Yu-Nung Su}
\affiliation{Institute of Astronomy and Astrophysics, Academia Sinica
11F of Astronomy-Mathematics Building
No.1, Sec. 4, Roosevelt Rd, Taipei 10617, Taiwan}

\author{Naomi Hirano}
\affiliation{Institute of Astronomy and Astrophysics, Academia Sinica
11F of Astronomy-Mathematics Building
No.1, Sec. 4, Roosevelt Rd, Taipei 10617, Taiwan}

\author{Motohide Tamura}
\affiliation{National Astronomical Observatory of Japan, National Institutes of Natural Sciences, Osawa, Mitaka, Tokyo 181-8588, Japan}
\affiliation{Department of Astronomy, Graduate School of Science, The University of Tokyo, 7-3-1 Hongo, Bunkyo-ku, Tokyo 113-0033, Japan}
\affiliation{Astrobiology Center, National Institutes of Natural Sciences, Osawa, Mitaka, Tokyo 181-8588, Japan}

\author{Masaaki Otsuka}
\affil{Okayama Observatory, Kyoto University, Kamogata, Asakuchi, Okayama 719-0232, Japan}

\author{Jennifer L. Karr}
\affiliation{Institute of Astronomy and Astrophysics, Academia Sinica
11F of Astronomy-Mathematics Building
No.1, Sec. 4, Roosevelt Rd, Taipei 10617, Taiwan}

\author{Tae-Soo Pyo}
\affiliation{Subaru Telescope, National Astronomical Observatory of Japan, National Institutes of Natural Sciences (NINS), 650 North A’ohoku Place, Hilo, HI 96720, USA}
\affiliation{School of Mathematical and Physical Science, SOKENDAI (The Graduate University for Advanced Studies), Hayama, Kanagawa 240-0193, Japan}

\begin{abstract}
A classical paradox in high-mass star formation is that powerful radiation pressure can halt accretion, preventing further growth of a central star.
Disk accretion has been proposed to solve this problem, but the disks and the accretion process in high-mass star formation are poorly understood.
We executed high-resolution ($R$=35,000-70,000) iSHELL spectroscopy in $K$-band for eleven high-mass protostars. 
%
%
Br-$\gamma$ emission was observed toward eight sources, and the line profiles for most of these sources are similar to those of low-mass PMS stars. Using an empirical relationship between the Br-$\gamma$ and accretion luminosities, we tentatively estimate disk accretion rates ranging from $\lesssim$10$^{-8}$ and $\sim$10$^{-4}$ $M_\sun$ yr$^{-1}$. These low-mass-accretion rates suggest that high-mass protostars gain more mass via episodic accretion as proposed for low-mass protostars.
%
%
Given the detection limits, CO overtone emission ($v$=2-0 and 3-1), likely associated with the inner disk region ($r \ll 100$ au), was found towards two sources.
This low-detection rate compared with Br-$\gamma$ emission is consistent with previous observations.
%
%
Ten out of the eleven sources show absorption at the $v$=0-2 ${\rm R(7)-R(14)}$ CO R-branch. Most of them are either blueshifted or redshifted, indicating that the absorption is associated with an outflow or an inflow with a velocity of up to $\sim50$ km s$^{-1}$. Our analysis indicates that the absorption layer is well thermalized (and therefore $n_{\mathrm H_2} \gtrsim 10^6$ cm$^{-3}$) at a single temperature of typically 100-200 K, and located within 200-600 au of the star.
\end{abstract}



\section{INTRODUCTION}
\label{sec:intro}
The formation process of high-mass stars (hereafter HM stars, $>8 M_\odot$) is poorly understood. The fundamental problem of HM star formation is that the powerful radiation pressure produced by the HM protostellar Objects (HMPOs) can considerably reduce mass accretion, preventing further mass growth. This is the so-called radiation pressure problem \citep{ka74,wo84}.
Theoretical work has suggested that a high-density disk can, by redistributing the radiation, reduce the radiation pressure on the accreting material in the midplane of the disk-envelope system, and therefore allow disk accretion to continue \citep{kr07,kr09,ku10,ku11}. 
Submillimeter-to-radio interferometric observations at high angular resolution have revealed a growing number of disks and disk ``candidates” (or torus-like envelopes, called ``toroids'') around HMPOs in recent years \citep{be14,jo15,il16,be16}.
This observing technique enables us to identify disks at $\gtrsim$100 au scales. However, it cannot trace
the mass accretion in the innermost disk region ($\ll$10 au from the star), which is essential for these stars to reach their final masses.

Near-IR spectroscopy is an alternative and powerful approach for investigating accretion in the inner disk region and/or from the disk to HM stars. This approach has been established for the evolution of low-mass young stellar objects (YSOs) and pre-main-sequence (PMS) stars \citep{na00,co10}, and it has been actively extended for HM star formation over the past decade. 
The emission features associated with HMPOs disappear when mass accretion terminates, and the spectra show \ion{H}{1} and \ion{He}{2} absorption associated with the photosphere of OB stars \citep[e.g.,][]{bi12}. 

For low-mass objects, Br $\gamma$ and other hydrogen recombination line emission at optical-to-IR wavelengths have been extensively observed to study mass accretion from the inner disk to the star.
According to the magnetospheric accretion paradigm, material from the inner disk accretes onto the protostar through the stellar magnetic channel.
The Br-$\gamma$ luminosity
is tightly correlated with the accretion luminosity and the disk accretion rates, therefore it 
has been widely used to estimate the disk accretion rate
\citep{mu98,na96,mo05}.
A few studies have extended the measured relationship between the Br-$\gamma$ luminosity and the disk accretion rate to HMPOs \citep{co13,po17}.
However, 
the Br $\gamma$ emission towards these objects could also originate from other components/regions such as a disk wind and/or an accretion disk \citep{bi05,ta14,ta16}.
Therefore, the Br $\gamma$ emission may not provide direct diagnostics of accretion onto HMPOs.

The CO ro-vibrational band emission may be another useful probe to study disk accretion onto HMPOs. The overtone emission at $\sim$2 \micron~($v=2-0$, $3-1$)
emitted from the hot ($T = 2500-5000 {\rm K}$) and dense ($n > 10^{11} {\rm cm^{-3}}$) gas, is likely
to trace the inner hot gaseous disk. 
The $v=2-0$ bandhead profile for some low-mass objects is associated with excess emission at the blueshifted side, which can be explained well by the broadening of individual lines due to disk rotation \citep[][for a review]{na00}. The luminosity of the bandhead emission is correlated to the Br-$\gamma$ emission \citep{co10}. Such studies of the bandhead profile and luminosity have been extended to intermediate protostars to HMPOs, and these show similar trends to low-mass objects \citep[e.g.,][]{da10,mu13,il13,il14,po17}.
Such a luminosity correlation supports the idea that, as for Br-$\gamma$, the CO emission is related to mass accretion.
However, the different line profiles and linewidths for CO and Br-$\gamma$ emission suggest that these two lines have different kinematic origins. 
Besides, the above observations have found that
the detection rates of CO are usually much lower than those of Br $\gamma$.

\citet{co13} conducted a low-resolution spectroscopic survey of massive YSOs and candidates following an infrared survey of the Galactic plane \citep[the Red MSX Source survey, hereafter RMS,][]{ho05,ur08}. This has been followed by detailed studies of selected samples using medium-to-high resolution spectroscopy \citep{il13,po17} and modeling work \citep{il18}, in particular for the CO overtone emission from the inner disk. In contrast, submillimeter-to-radio interferometry, a conventional and powerful tool for studying circumstellar disks and disk candidates associated with HM stars, has been extensively applied to those objects which are bright at these wavelengths \citep[e.g.,][for a review]{be16}.
The RMS studies may not statistically represent the properties of these millimeter sources, which tend to be heavily embedded at infrared wavelengths \citep[e.g.,][see also Section 2]{ta12}.

Here we present a $K$-band high-resolution spectroscopic survey of eleven HM 
protostars associated with a circumstellar disk or disk candidate identified at submillimeter-to-radio wavelengths. In Section \ref{sec:sample} we describe how we selected the samples.
In Section \ref{sec:obs} we describe the observations. 
The observed line and properties are presented in Section \ref{sec:result}, and the detailed analysis and discussion are given in Section \ref{sec:discussion}. 
Section \ref{sec:summary} summarizes the conclusions of this paper.


\section{Sample Selection}
\label{sec:sample}
We selected eleven high-mass protostars (HMPOs) from \citet{be16}, who listed 41 HMPOs associated with a disk or a disk candidate observed at submillimeter-to-radio wavelengths.
These are listed in Table \ref{tab:tar} with their stellar and disk parameters, and the presence/absence of an associated UC/HC HII region.
We selected sources with a bright near-IR counterpart (either as a reflection/emission nebula or a point source) by examining the 2MASS images in order to conduct the infrared spectroscopic observations.
Furthermore, we selected sources that were relatively isolated in the 2MASS images to avoid contamination or confusion from other sources. 

\begin{deluxetable*}{ccccccccccccc}
\newcounter{tabref}
\newenvironment{tabref}[1][]{\refstepcounter{tabref} (\thetabref)#1}{\noindent}
\tabletypesize{\scriptsize}
\tablecaption{Targets}
\tablehead{ 	
\colhead{Name}		
& \colhead{R.A.}
& \colhead{Dec}	
& \colhead{$L_{\rm bol}$}	
& \colhead{$d$}
& \colhead{$V_{\rm lsr}$}
& \colhead{$M_{\rm \star, L}$}
& \colhead{$M_{\rm \star, c}$}
& \colhead{$R_{\rm disk}$}	
& \colhead{\ion{H}{2}}	
& \colhead{$\theta_{\rm inc}$}
& \colhead{exp.}
& \colhead{Ref.}	
\\
\colhead{}		
& \colhead{hms}
& \colhead{dms}	
& \colhead{$L_\odot$}		
& \colhead{kpc}
& \colhead{${\rm km~s^{-1}}$}
& \colhead{$M_\odot$}
& \colhead{$M_\odot$}
& \colhead{a.u.}
& \colhead{}
& \colhead{deg.}
& \colhead{min.}
& \colhead{}
}
\startdata 
W3 IRS 5			&  02h25m40.8s	& +62d05m52.6s 	& 2$\times$10$^5$	& 2.0   & $^{\ref{wa13}}$-39.5	& - &	35	& 2000		& $^{\ref{cl94}}$HC \ion{H}{2}	& - & 30 & \ref{be16},\ref{cl94},\ref{wa13}\\
AFGL 490			&  03h27m38.8s	& +58d47m00.1s 	& 2$\times$10$^3$	& 1.0	& $^{\ref{sc06}}$-13.4	& 8-10	& 7	& 1600 & $^{\ref{sc02}}$HC \ion{H}{2}	& $^{\ref{sc06}}$30 & 2 & \ref{be16},\ref{co13},\ref{sc02},\ref{sc06}\\
IRAS 04579+4703	&  05h01m39.7s	& +47d07m21.9s 	& 4$\times$10$^3$	& 2.5	    & $^{\ref{xu12}}$-17.0 & 7	& 8.5		& 5000	& $^{\ref{sa08}}$Perhaps UC \ion{H}{2}	& -& 35 & \ref{be16},\ref{sa08},\ref{xu12}\\
S255IR SMA1		&  06h12m54.0s	& +17d59m23.0s 	& 2$\times$10$^4$	& 1.59	    & $^{\ref{zi12}}$4.4   & -	& 14	& 1850	& $^{\ref{oj11},\ref{sn86}}$UC \ion{H}{2}	& $^{\ref{li20}}$60	&26	&\ref{be16},\ref{co13},\ref{li20},\ref{oj11},\ref{sn86},\ref{zi12},\ref{zi15}\\
S255IR SMA2		&  06h12m53.8s	& +17d59m23.0s 	& 2$\times$10$^4$	& 1.59	    & $^{\ref{zi12}}$9.2   & -	& 14	& 1850	& $^{\ref{oj11},\ref{sn86}}$No	& -	& 20	& \ref{be16},\ref{co13},\ref{li20},\ref{oj11},\ref{sn86},\ref{zi12},\ref{zi15}\\
W33A-MM1 Main	&  18h14m39.5s	& $-$17d52m00.5s 	& 1$\times$10$^5$	& 3.8	& $^{\ref{ga10}}$38.5  & -	& 25 & 1900		& $^{\ref{iz18},\ref{va05}}$HC \ion{H}{2}	& $^{\ref{de10}}$50	& 10	& \ref{be16},\ref{iz18},\ref{va05},\ref{de10},\ref{ga10},\ref{po17},\ref{ga10}\\
IRAS 18151-1208	&  18h17m58.1s	& $-$12d07m24.8s 	& 2$\times$10$^4$	& 3.0	& $^{\ref{sr02}}$32.8  & 15		& 14	& 5000	& $^{\ref{be05}}$No	& $^{\ref{fa11}}$60	& 10	& \ref{be16},\ref{co13},\ref{po17},\ref{fa11},\ref{be05},\ref{sr02}\\
G29.96-0.02		&  18h46m04.0s	& $-$02d39m21.7s 	& 8$\times$10$^5$	& 6.2	&  $^{\ref{be11}}$92.9    & 33		& 56	& 4000	& $^{\ref{be11}}$UC \ion{H}{2}	& -	& 35	& \ref{be16},\ref{be11}\\
AFGL 2591 VLA3	&  20h29m24.9s	& +40d11m19.4s 	& 2$\times$10$^4$	& 1.0	    &  $^{\ref{wa12}}$-5.4  & 16	& 14	& 400		& $^{\ref{jo13}}$HC \ion{H}{2}	&  $^{\ref{de06}}$26-38	& 2	& \ref{be16},\ref{co13},\ref{po17},\ref{jo13},\ref{de06},\ref{wa12}\\
NGC 7538 IRS 1	&  23h13m45.3s	& +61d28m11.7s 	& 8$\times$10$^4$	& 2.7	    & $^{\ref{be12}}$-57.3 & 30	& 23 & 1000		& $^{\ref{mo14}}$Perhaps UC \ion{H}{2}	& -	& 4	& \ref{be16},\ref{po17},\ref{be12},\ref{mo14}\\
IRAS 23151+5912	&  23h17m21.4s	& +59d28m49.1s 	& 1$\times$10$^5$	& 5.7	    & $^{\ref{ro14}}$-61.4 &  8	& 25	& 2150		& $^{\ref{ro14}}$HC \ion{H}{2} 	& -	& 20	& \ref{be16},\ref{co13},\ref{po17},\ref{ro14}
\enddata
Col. (4) Bolometric luminosity \citep{be16}.
Col. (5) Distance \citep{be16}.
Col. (6) Systemic velocity of the targets from literature. The prefix number indicates the sources of references (see Col. (13) below).
Col. (7) Stellar mass estimated from the Lymann-continuum photons \citep{be16}.
Col. (8) Stellar mass derived using $L_{\rm bol}$, performing comparisons with cluster models and removing contributions from the star with lower masses \citep{be16}. The derived masses using this method are in good agreement with those obtained using the other method, but this method is applicable to sources without an associated \ion{H}{2} region.
Col. (9) Size of an associated disk or disk candidate measured using millimeter interferometry \citep{be16}.
Col. (10) Type of associated \ion{H}{2} region. The prefix number indicates the sources of references (see Col. (13) below).
Col. (11) Inclination angle (0 degrees for face-on case) of the disk or the disk candidate measured from millimeter/submillimeter observations. The prefix number indicates the sources of references (see Col. (13) below).
Col. (12) Total on-source exposure.\\
Col. (13) Reference:
\begin{tabref} \citet{be16};\label{be16} \end{tabref}
\begin{tabref} \citet{cl94};\label{cl94} \end{tabref}
\begin{tabref} \citet{wa13};\label{wa13} \end{tabref}
\begin{tabref} \citet{co13};\label{co13} \end{tabref}
\begin{tabref} \citet{sc02};\label{sc02} \end{tabref}
\begin{tabref} \citet{sc06};\label{sc06} \end{tabref}
\begin{tabref} \citet{sa08};\label{sa08} \end{tabref}
\begin{tabref} \citet{xu12};\label{xu12} \end{tabref}
\begin{tabref} \citet{li20};\label{li20} \end{tabref}
\begin{tabref} \citet{oj11};\label{oj11} \end{tabref}
\begin{tabref} \citet{sn86};\label{sn86} \end{tabref}
\begin{tabref} \citet{zi12};\label{zi12} \end{tabref}
\begin{tabref} \citet{zi15};\label{zi15} \end{tabref}
\begin{tabref} \citet{iz18};\label{iz18} \end{tabref}
\begin{tabref} \citet{va05};\label{va05} \end{tabref}
\begin{tabref} \citet{de10};\label{de10} \end{tabref}
\begin{tabref} \citet{ga10};\label{ga10} \end{tabref}
\begin{tabref} \citet{po17};\label{po17} \end{tabref}
\begin{tabref} \citet{ga10};\label{ga10} \end{tabref}
\begin{tabref} \citet{jo13};\label{jo13} \end{tabref}
\begin{tabref} \citet{fa11};\label{fa11} \end{tabref}
\begin{tabref} \citet{be05};\label{be05} \end{tabref}
\begin{tabref} \citet{sr02};\label{sr02} \end{tabref}
\begin{tabref} \citet{be11};\label{be11} \end{tabref}
\begin{tabref} \citet{de06};\label{de06} \end{tabref}
\begin{tabref} \citet{wa12};\label{wa12} \end{tabref}
\begin{tabref} \citet{be12};\label{be12} \end{tabref}
\begin{tabref} \citet{mo14};\label{mo14} \end{tabref}
\begin{tabref} \citet{ro14}\label{ro14} \end{tabref}

\label{tab:tar}
\end{deluxetable*} 

To investigate the near-infrared excess associated with warm circumstellar dust, we plotted a $J-H$ vs. $H-K$ color-color diagram, which is often used for low-mass PMS stars \citep[e.g.,][]{Stahler05} (Figure \ref{fig:ccd}). For our sample we used the 2MASS $JHK$ magnitudes (see Appendix \ref{sec:appendix:2mass}). 
For reference, we also plotted the data from \citet{co13}, who obtained low-resolution near-IR spectra for $\sim$130 massive YSOs. The diagram shows that most of our sample show significant $K$-band excesses compared with main-sequence stars and giants (even including reddening) as does the Cooper sample. This trend suggests 
that our sources are associated with significant amounts of 
warm dust, potentially associated with the inner disk region.

\begin{figure}
\includegraphics[width=0.5\textwidth]{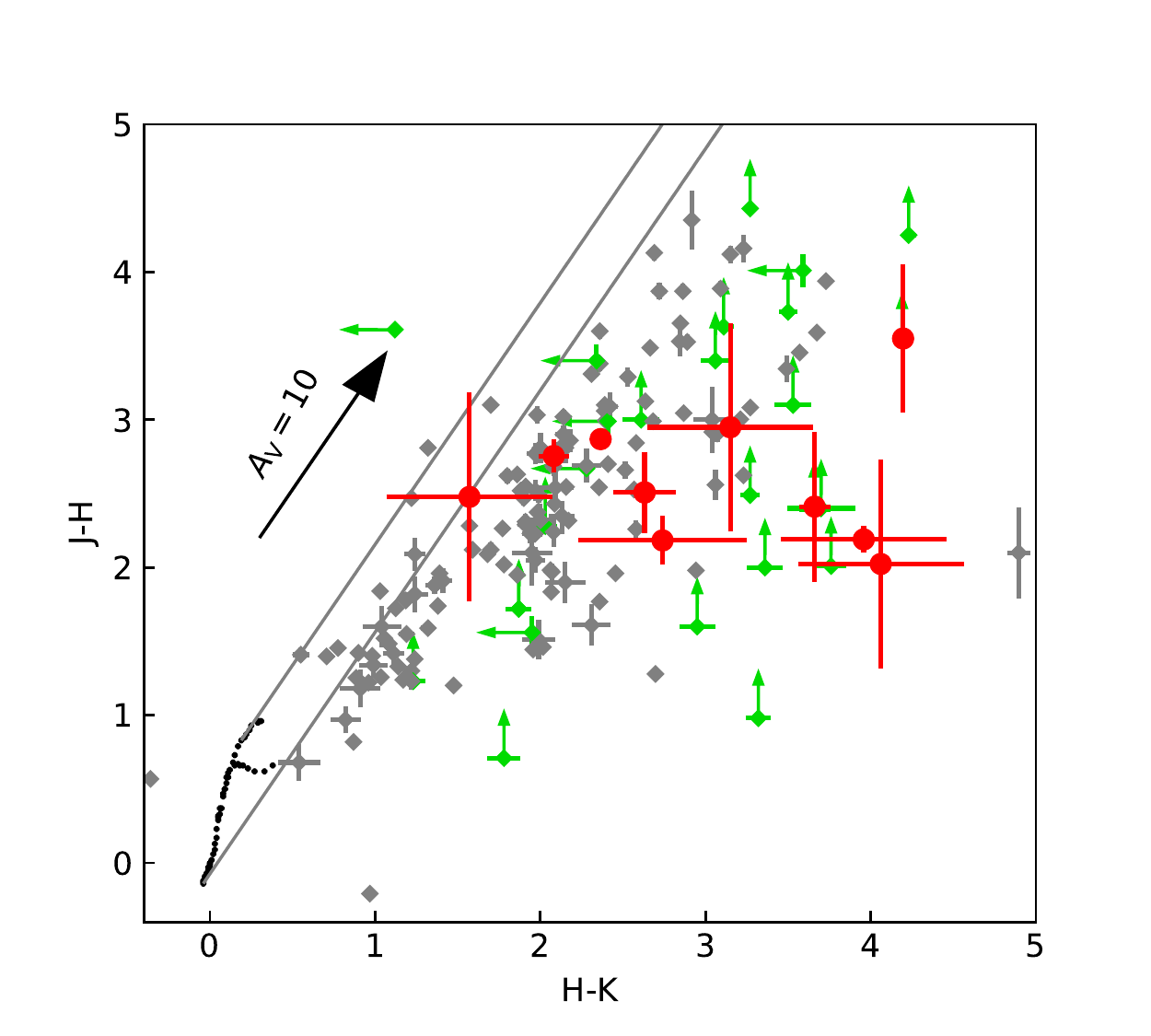}
\caption{$JHK$ color-color diagrams for HMPOs, young stars and main-sequence stars. The data points for our sources (see Table \ref{tab:2mass} for $JHK$ magnitudes) are shown in red. 
We assume an uncertainty of 0.5 magnitudes for those for which the 2MASS catalogue does not provide an uncertainty (Appendix A).
As S255IR SMA1/SMA2 are not resolved in the 2MASS images, these two objects share the same color/extinction in this work. 
The grey and green (as lower/upper limit) points are HM stars identified by \citet{co13} based on the RMS survey and follow-up near-IR spectroscopy. The black dots in the bottom-left of the panel are main-sequence and giant stars without extinction; the gray lines show the expected reddened colors from these stars. 
}
\label{fig:ccd}
\end{figure}

Our target objects have a variety of stellar masses ($M_\star=7-56~M_\odot$), and bolometric luminosities ($L_{\rm bol} =10^3 - 10^6~L_\odot$). The luminosity is considered to be a sum of the stellar and accretion luminosities. Four to eight sources are associated with either a hypercompact \ion{H}{2} (HC) or ultracompact \ion{H}{2} (UC) region due to illumination of the circumstellar gas by stellar UV radiation.
Such a variety in luminosity and the presence/absence of the associated H II region allows us to discuss the target features, and therefore disk evolution, for a variety of evolutionary phases as well as a variety of masses.


\section{Observations and Data Reduction}
\label{sec:obs}
We observed the above HMPOs using iSHELL (\citealt{va03,cu04,ra16}) on the NASA Infrared Telescope Facility (IRTF) \footnote{Visiting Astronomer at the Infrared Telescope Facility, which is operated by the University of Hawaii under contract 80HQTR19D0030 with the National Aeronautics and Space Administration.} on 2017 September 20th and 22nd, at a 0\farcs6-0\farcs8 seeing. The program ID is 2017B032.
The pointing position was chosen toward the emission peak at near-infrared during the observation.
The wavelength setting was designed to cover 2.024 to 2.334 $\mu$m, cross-dispersed in 35 echelle orders with a spectral resolution $R$ of up to 70,000.
For this paper we focus our study on the Br $\gamma$ and CO overtone transitions, i.e., the two major transitions discussed in Section \ref{sec:intro}. 

Each source was observed with one or two object-sky-sky-object sequences.
The total on-source time for each source is listed in Table \ref{tab:tar}.
During the observations, the signal-to-noise (S/N) ratio was simultaneously estimated for each source.
Then, based on the S/N ratio, we selected a slit width of $0\farcs375\times5\arcsec$ or $0\farcs75\times5\arcsec$ and decided the integration time in order to achieve a reasonable S/N ratio for each source.
These slit widths result in spectral resolutions of $R\sim70,000$ and $R\sim$35,000, corresponding to velocity resolutions of 4.3 and 8.6 ${\rm km~s^{-1}}$, respectively.
The exposure times were $2-35$ min depending on the source brightness.
For the telluric calibration, we observed A-type stars every $\sim2$ hours.

The data were reduced and analyzed using PyRAF
(\citealt{pyraf}),
and several python packages, scipy, numpy, and matplotlib.
We applied the standard data reduction process including the sky subtraction from the object-sky-sky-object sequence, flat fielding and wavelength calibration.
To calibrate the wavelength in each echelle order, we first derived a 1-D spectrum combined all the orders using a few well-known bright Thorium-Argon lines.
We later identified more Thorium-Argon lines from the 1-D wavelength-calibrated spectrum in each order and used them to obtain the 2-D images (with the space-wavelength dimensions) by solving the 3rd-order Chebyshev equation.
By comparing the observed telluric absorption lines with those computed by ATRAN \citep{lo92},
we estimate the error of the wavelength calibration to be lower than $\sim$3 km s$^{-1}$
Although we subtracted sky frames, we found residual background due to imperfect sky subtraction and/or a diffuse reflection nebula associated with the target. We measured it near the edges of the slit (emission free region) and subtracted it for each wavelength pixel.
We then extracted the 1-D spectra by integrating a limited spatial region to maximize signal-to-noise.
Telluric corrections were applied by dividing the normalized spectrum of the standard star; 
the telluric features from the standard stars were scaled by a factor to fit the target spectrum as the airmass can change on a time scale of hours.


\section{Results}
\label{sec:result}

In Sections 4.1-4.3 we describe the results for the Br-$\gamma$ emission/absorption, CO $v$=2--0 bandhead emission and CO $v$=0--2 absorption, respectively.


\subsection{B\lowercase{r}$\gamma$ Emission/Absorption}
\label{sec:result:Br}
\begin{figure*}
\includegraphics[width=0.95\textwidth]{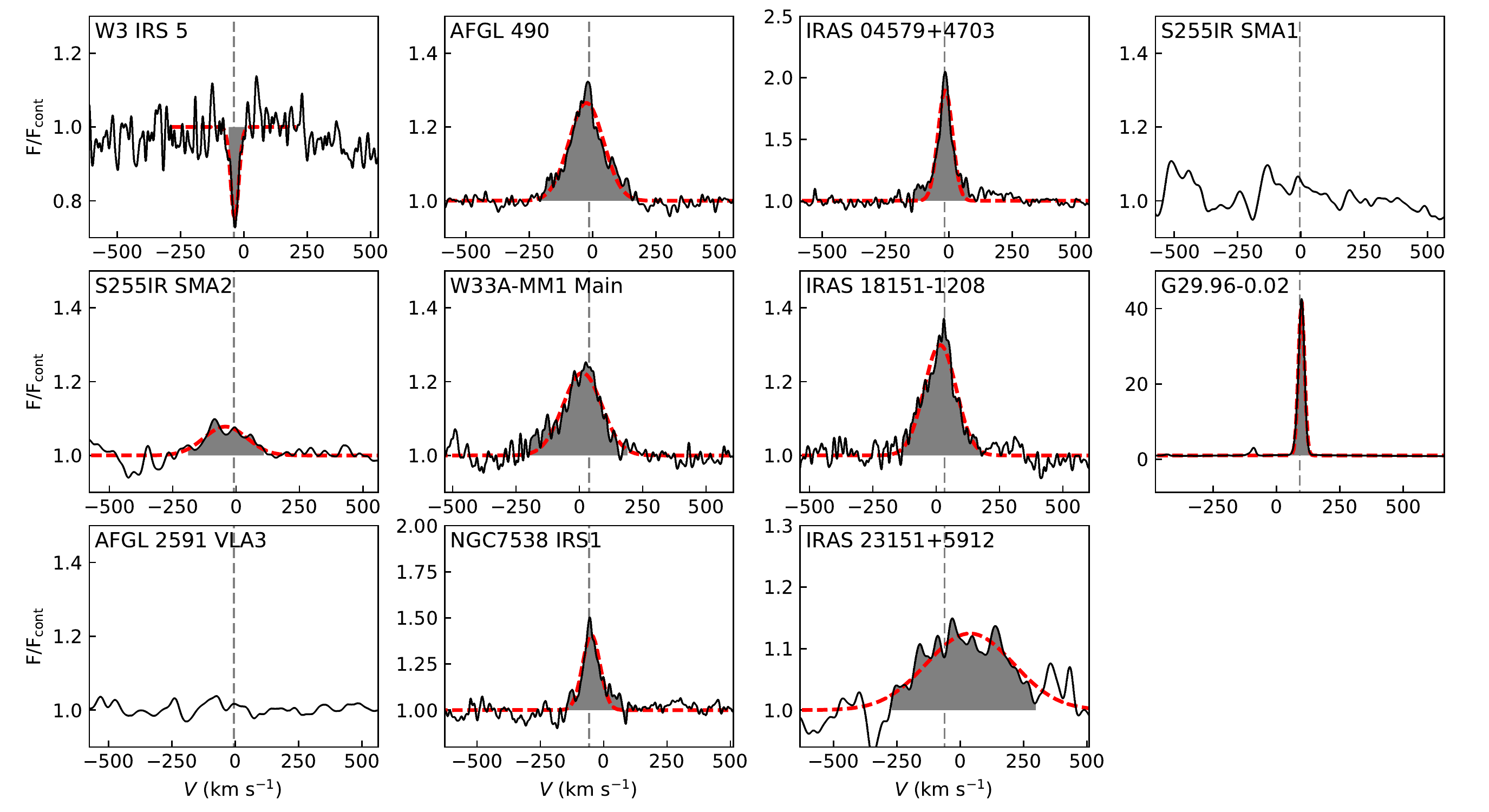}
\caption{Br $\gamma$ line profiles toward the eleven massive protostars. 
The dashed line in each panel shows the velocity of the source based on millimeter observations (see Table \ref{tab:tar} for references). The gray regions show the ranges used for measuring the equivalent widths (Table \ref{tab:br}).
}
\label{fig:br}
\end{figure*}

Figure \ref{fig:br} shows the Br $\gamma$ line profiles for eleven sources.
In Table \ref{tab:br} we list the equivalent widths of the emission/absorption lines, and by Gaussian fitting, the measured FWHMs.
Line emission is observed toward eight sources over all the categories of \ion{H}{2} region associations,
including ``No \ion{H}{2} region'' (IRAS 18151-1108 and S255 IR SMA2), HC \ion{H}{2} region (IRAS 04579+4703 and IRAS 23151+5912), and UC \ion{H}{2} region (AFGL 490, W33A-MM1 Main, G29.96-0.02, and NGC 7538 IRS1). All but G29.96-0.02  show a triangle-like profile with a full width half maximum (FWHM) of $\sim70-400~{\rm km~s^{-1}}$, while G29.96-0.02 shows a significantly narrower linewidth (FWHM$\sim 28~{\rm km~s^{-1}}$). Out of the remaining three sources, W3 IRS 5 shows an absorption line with a FWHM of $\sim 37~{\rm km~s^{-1}}$.

The measured equivalent widths of the emission lines range from $\sim-1.1$ to $-5.4$ \AA~ except for G29.96-0.02, which has an equivalent width larger by a factor of  20--80 ($\sim90$ \AA). 
For non-detected sources we estimated an upper limit of the equivalent width with a S/N$<3$ assuming a line width of 50 km s$^{-1}$.
The equivalent widths and the upper limits reported here are broadly consistent with those measured by previous observations at low resolutions (Appendix \ref{app:br}).


\tabletypesize{\scriptsize}
\begin{deluxetable*}{ccccccccc}
\tabletypesize{\footnotesize}
\tablecaption{Properties of $\rm Br\gamma$ Emission/Absorption}
\tablehead{ 	
\colhead{(1) Name}
& \colhead{(2) EW}	
& \colhead{(3) $\Delta V$}
& \colhead{(4) $A_{V} (JH)$}
& \colhead{(5) $L_{\rm Br\gamma}$}
& \colhead{(6) $L_{\rm acc}$}
& \colhead{(7) $\dot{M}_{\rm acc,model}$}
& \colhead{(9) $\dot{M}_{\rm acc,cluster}$}
\\
\colhead{}		
& \colhead{\AA}
& \colhead{${\rm km~s^{-1}}$}	
& \colhead{}
& \colhead{$L_\odot$}
& \colhead{$L_\odot$}
& \colhead{$M_\odot$ yr$^{-1}$}
& \colhead{$M_\odot$ yr$^{-1}$}
}
\startdata 
W3 IRS 5        & 0.66$\pm$0.03        & 36.6$\pm$2.1  & 19.4$\pm$7.2  & -     & -     & -     & -\\
AFGL 490        & -3.39$\pm$0.02 & 160.8$\pm$1.6 & 27.9$\pm$0.5  & 0.21$\pm$0.01   & $847 \pm 28$  & $\left(1.26 \pm 0.04\right) \times 10^{-5}$   & $\left(1.33 \pm 0.04\right) \times 10^{-5}$\\
IRAS 04579+4703 & -5.37$\pm$0.05 & 67.9$\pm$0.9  & 21.0$\pm$1.7  & 0.011$\pm$0.001       & $60 \pm 7$    & $\left(8.2 \pm 1.0\right) \times 10^{-7}$     & $\left(8.9 \pm 1.1\right) \times 10^{-7}$\\
S255IR SMA1     & $<$0.06       & -     & 24.3$\pm$2.8  & $<$0.0001      & $<$0.8        & $<8.7 \times 10^{-9}$    & $<9.7 \times 10^{-9}$\\
S255IR SMA2     & -1.119$\pm$0.004 & 200.3$\pm$7.2 & 24.3$\pm$2.8  & 0.0019$\pm$0.0004       & $11.7 \pm 2.3$        & $\left(1.25 \pm 0.25\right) \times 10^{-7}$   & $\left(1.39 \pm 0.27\right) \times 10^{-7}$\\
W33A-MM1 Main   & -3.17$\pm$0.05 & 180.2$\pm$2.9 & 21.0$\pm$0.9  & 0.066$\pm$0.005       & $300 \pm 20$  & $\left(2.36 \pm 0.16\right) \times 10^{-6}$   & $\left(2.77 \pm 0.18\right) \times 10^{-6}$\\
IRAS 18151-1208 & -3.43$\pm$0.04 & 148.9$\pm$2.0 & 23.3$\pm$5.2  & 0.051$\pm$0.020       & $\left(2.3 \pm 0.9\right) \times 10^{2}$      & $\left(2.5 \pm 0.9\right) \times 10^{-6}$     & $\left(2.8 \pm 1.0\right) \times 10^{-6}$\\
G29.96-0.02     & -89.76$\pm$0.05        & 28.1$\pm$0.1  & 26.8$\pm$1.2  & 9.6$\pm$0.9   &  -\tablenotemark{a}   & -\tablenotemark{a}   & -\tablenotemark{a} \\
AFGL 2591 VLA3  & $<$0.03       & -     & 34.9$\pm$5.1  & $<$0.001      & $<$9.1        & $<9.7 \times 10^{-8}$    & $<1.1 \times 10^{-7}$\\
NGC 7538 IRS1    & -2.83$\pm$0.05 & 80.9$\pm$1.7  & 28.8$\pm$7.2  & 0.1$\pm$0.1   & $\left(4.5 \pm 2.3\right) \times 10^{2}$      & $\left(3.8 \pm 1.9\right) \times 10^{-6}$     & $\left(4.3 \pm 2.2\right) \times 10^{-6}$\\
IRAS 23151+5912 & -3.52$\pm$0.11 & 404.6$\pm$11.1        & 24.0$\pm$7.2  & 0.3$\pm$0.2   & $\left(1.3 \pm 0.6\right) \times 10^{3}$      & $\left(1.0 \pm 0.5\right) \times 10^{-5}$     & $\left(1.2 \pm 0.6\right) \times 10^{-5}$
\enddata
\tablecomments{
Col. (2) Equivalent widths. 
Col. (3) Full width half-maximum velocity obtained from the Gaussian fitting. 
Col. (7) Disk accretion rate estimated using $\dot{M}_{\rm acc}=\frac{L_{\rm acc}R_{\star}}{GM_{\star}}$ with $M_{\star}$ as a function of $L_{\rm bol}$ from the model in \citet{da11}. 
Col. (8) Disk accretion rate estimated using $\dot{M}_{\rm acc}=\frac{L_{\rm acc}R_{\star}}{GM_{\star}}$ with $M_{\rm star}=M_{\rm \star,cluster}$
}

\tablenotetext{a}{The emission is probably associated with a UC \ion{H}{2} region, which is not related to the parameters for mass accretion. See text for details.
}
\label{tab:br}
\end{deluxetable*} 


\subsection{CO Bandhead Emission}
\label{sec:result:coemission}
\begin{figure*}
\includegraphics[width=1.02\textwidth]{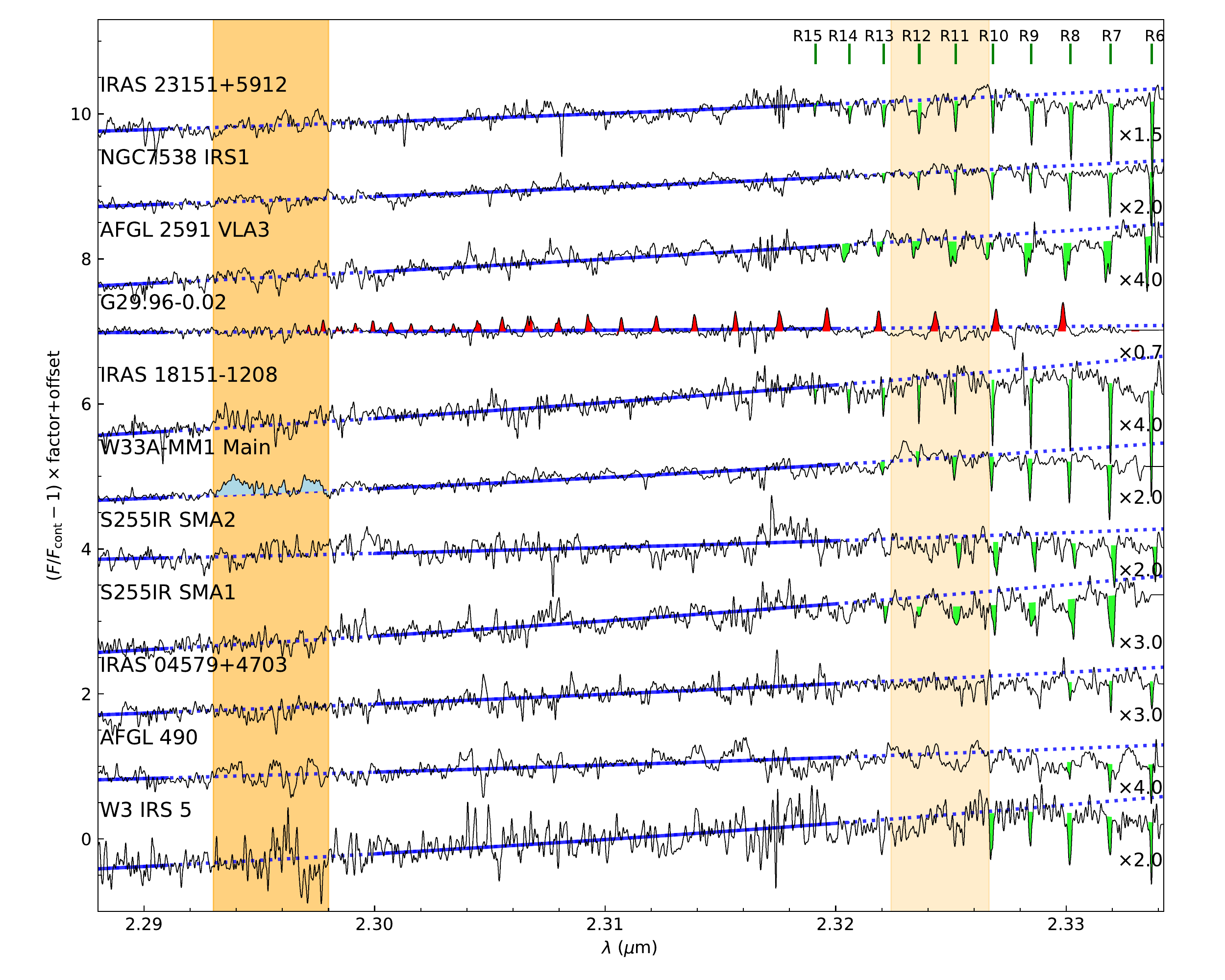}
\caption{Spectra of CO $v=2-0$, CO $v=3-1$, and the Pfund series. 
Each spectrum is normalized to its continuum flux at $2.29-2.33~\mu$m and scaled by the factor labeled at the right hand side. The spectrum in between each source is offset by a factor of one.
The wavelengths are shifted according to the target's systematic velocity from the submm/mm observations in the literature. 
The colored areas show the detected Pfund emission (red), CO $v=2-0$ absorption (green) and CO $v=2-0$ bandhead emission (light blue). The orange areas highlight the wavelengths of the CO $v=2-0$ (dark orange) and $v=3-1$ (light orange) bandhead emission we used to measure the EWs, respectively.}

\label{fig:co}
\end{figure*}

Figure \ref{fig:co} shows the spectra at 2.288--2.334 \micron, which covers the CO $v=2-0$ and $3-1$ bandhead emission.
Both bandheads are seen in emission in W33A MM1 Main, and the $v=2-0$ bandhead is marginally detected in IRAS 18151-1208.

To measure the equivalent width of the $v=2-0$ bandhead emission, we calculated the continuum baseline by employing a 3rd-order polynomial fit in the line-free regions of $2.190-2.291~\mu$m and $2.300-2.320~\mu$m and subtracted it.
We then integrated the spectrum over $2.293-2.298~\mu$m.
The equivalent widths measured for W33A MM1 Main and IRAS 18151-1208 are $-2.33\pm0.69$ and $-0.67\pm0.55$ \AA, respectively.
For the remaining sources, we derived a 3-$\sigma$ upper limit for the CO $v=2-0$ emission of $-1.0$ to $-4.4$ \AA. Near-IR spectroscopy at low-to-medium resolutions has already been done for most of the sources, and the equivalent widths and the upper limits we measured are broadly consistent with the previous observations (Appendix C).

We do not derive the equivalent width of the $v=3-1$ bandhead emission observed toward W33A MM1 Main
for the following reasons: (1) narrow $v=2-0$ absorption lines cover the same wavelength range (see Figure \ref{fig:co} and section \ref{sec:co abs}); and (2) our spectra do not cover the continuum at the longer wavelengths, making the continuum fitting inaccurate.




\subsection{CO $\lowercase{v}=0-2$ Absorption}
\label{sec:co abs}

\begin{figure*}
\includegraphics[width=0.96\textwidth]{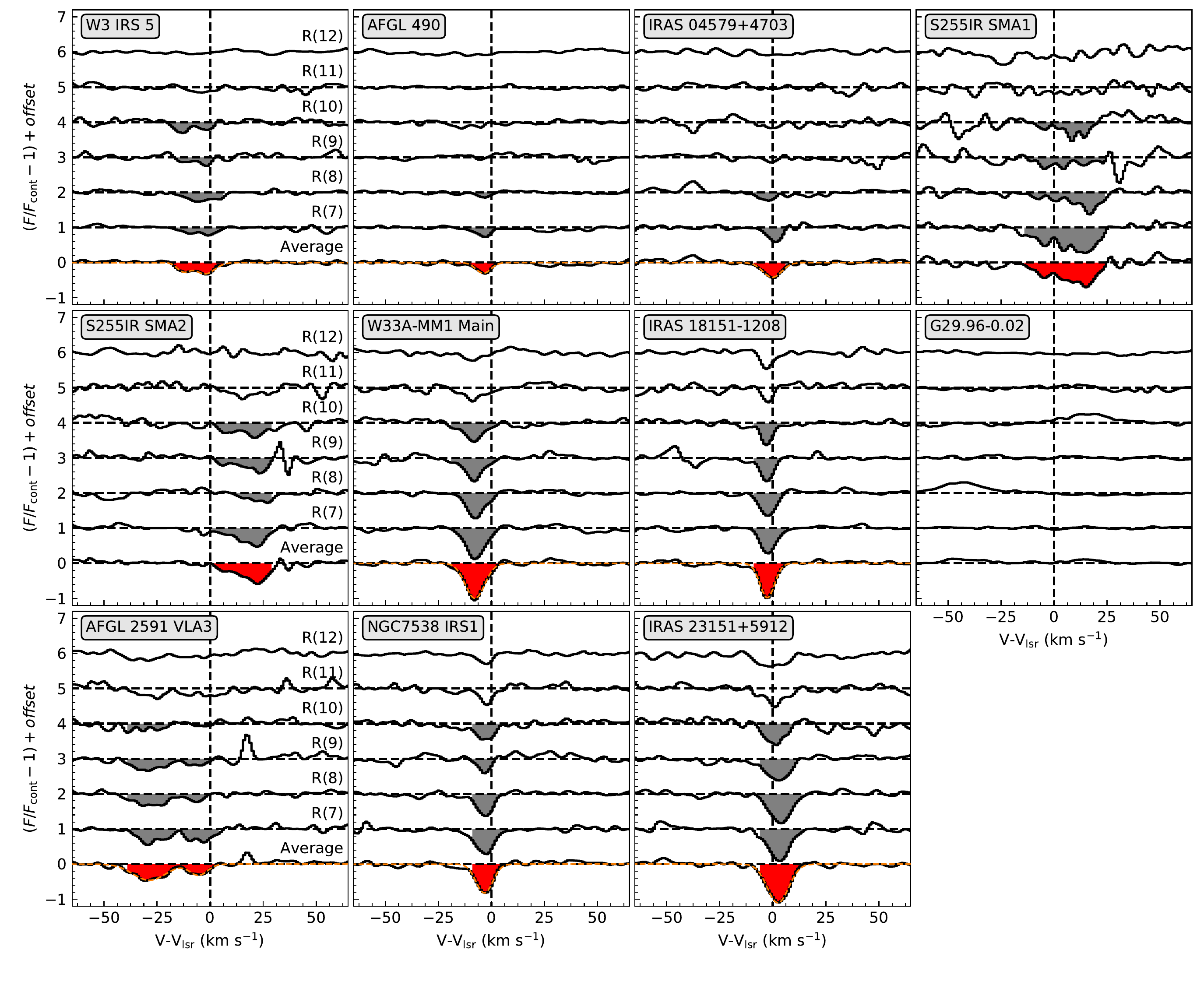}
\caption{
CO R(7)-R(12) line profiles for all the sources. 
Each line profile is normalized to the continuum level. The bottom profile in each panel shows the averaged line profiles using the R(7)-R(8) transitions for AFGL 490 and IRAS 04579+4703, and R(7)-R(10) for the remaining sources. For all but S255IRS SMA1 and SMA2, we show the Gaussian fit at the bottom of each panel (with an orange dashed curve; see text for details).}

\label{fig:stack}
\end{figure*}

In Figure \ref{fig:co} all  sources except G29.96-0.02 show narrow CO $v=2-0$ absorption lines at 2.32--2.33 \micron.
In Figure \ref{fig:stack} we show the $R$(7)--$R$(12) profiles in detail. For these profiles the velocities are shown with respect to the source velocity measured using the millimeter observations. Table \ref{tab:co} shows their equivalent widths.

\begin{deluxetable*}{cccccccccccc}
\label{tab:co}
\tabletypesize{\footnotesize}
\tablecaption{Equivalent Widths of CO $v=2-0$ Absorption (\AA)
}
\tablehead{ 	
\colhead{Name}		
& \colhead{R6 }
& \colhead{R7 }	
& \colhead{R8 }	
& \colhead{R9 }	
& \colhead{R10}	
& \colhead{R11 }	
& \colhead{R12 }	
& \colhead{R13 }	
& \colhead{R14 }	
& \colhead{R15 }
& \colhead{rms}
}
\startdata 
W3 IRS 5        & 0.38  & 0.38  & 0.53  & 0.28  & 0.44  & -     & -     & -     & -     & -     & 0.10  \\
AFGL 490        & 0.13  & 0.11  & 0.07  & -     & -     & -     & -     & -     & -     & -     & 0.09  \\
IRAS 04579+4703 & 0.12  & 0.12  & 0.08  & -     & -     & -     & -     & -     & -     & -     & 0.03  \\
S255IR SMA1     &  -  & 0.49  & 0.36  & 0.25  & 0.20  & 0.21  & 0.07  & 0.09  & -     & -     & 0.11  \\
S255IR SMA2     &  0.26 & 0.42  & 0.20  & 0.27  & 0.38  & 0.25  & -     & -     & -     & -    & 0.09  \\
W33A-MM1 Main   & -      & 0.42  & 0.28  & 0.35  & 0.28  & 0.20  & 0.12  & 0.13  & -     & -     & 0.05  \\
IRAS 18151-1208 & 0.32  & 0.25  & 0.20  & 0.20  & 0.20  & 0.07  & 0.10  & 0.09  & 0.07  & 0.04  & 0.03  \\
G29.96-0.02     &  -    & -     & -     & -     & -     & -     & -     & -     & -     & -     & -     \\
AFGL 2591 VLA3  & 0.24  & 0.34  & 0.28  & 0.24  & 0.10  & 0.23  & 0.12  & 0.09  & 0.16  & -     & 0.07  \\
NGC7538 IRS1    & 0.33  & 0.30  & 0.23  & 0.09  & 0.17  & 0.12  & 0.10  & 0.07  & 0.01  & -     & 0.03  \\
IRAS 23151+5912 & 0.64  & 0.59  & 0.60  & 0.46  & 0.23  & 0.30  & 0.35  & 0.23  & 0.16  & 0.02  & 0.07  
\enddata
\end{deluxetable*} 

To increase signal-to-noise ratios, we averaged the $R$(7)-$R$(10) profiles (i.e.\ those with the largest absorption) for all the sources but AFGL 490 and IRAS 04579+4703. These line profiles are shown at the bottom of each panel. For the latter two sources we use $R$(7) and $R$(8) only for the averaged spectra due to relatively low signal-to-noise of the upper transitions.

The absorption line profiles show a rich variety between the sources. Given the systemic velocity from sub/millimiter observations (Table {\ref{tab:tar}}), we found that, for S255IR SMA1 the absorption is deepest at $\Delta V$$\sim$15 km s$^{-1}$, with blueshifted wing absorption toward $\Delta V$$\sim$$-$15 km s$^{-1}$. S255IR SMA2 shows a similar line profile but with an absorption maximum at $\Delta V$$\sim$20 km s$^{-1}$ and a blueshifted wing down to $\Delta V$$\sim$0 km s$^{-1}$.

AFGL 2591 VLA3 shows two separated absorption components, with absorption maxima at $\Delta V\sim$$-30$ and $-5$ km s$^{-1}$, respectively. We measure FWHM widths of 18 and 10 km s$^{-1}$, respectively, using Gaussian fitting. 
The line profiles observed toward W3 IRS 5 can be also attributed to two components with Gaussian profiles, with peak velocities of $-$13 and $-$3 km s$^{-1}$, respectively. The Gaussian fitting indicates that the individual absorption components are not clearly resolved at the given velocity resolution of 8.6 km s$^{-1}$.

The remaining sources show a single absorption component line with a Gaussian profile.
The absorption maximum is blueshifted by 3-8 km s$^{-1}$ for AFGL 490, W33-MM1 Main, IRAS 18151-1208 and NGC 7538 IRS1; and redshifted by $\sim$ 2 km s$^{-1}$ for IRAS 23151+5912. The absorption component for IRAS 04579+4703 does not show any clear evidence for a blueshift or redshift. Those for AFGL 490, IRAS 18151-1208 and IRAS 23151+5912 have a Gaussian-fitted FWHM width of 8-12 km $^{-1}$. The remaining profiles are not resolved at the given velocity resolutions of 4.3 and 8.6 km s$^{-1}$.


\section{Discussion}
\label{sec:discussion}

In Section 5.1 we briefly describe the extinction corrections for line and band luminosities. In Section 5.2 we discuss the implications of the Br-$\gamma$ emission for HM star formation. In Section 5.3 we discuss the low detection rate of CO overtone emission. In Section 5.4 we discuss the nature of the CO $v$=0-2 absorption.

\subsection{Extinction Correction}
\label{sec:av}

We estimate extinction toward each source using the method adopted by \citet{co13} as described below. We calculate
two extinctions; $A_{V} (JH)$ and $A_{V} (HK)$, using the $JHK$ magnitude with the following equation:
\begin{equation}
A_{ V}=\frac{m_1-m_2+c_{\rm int}}{0.55^{1.75}(\lambda^{-1.75}_1-\lambda^{-1.75}_2)}
\end{equation}
where $m_1$ and $m_2$ are the magnitudes, $\lambda_1$ and $\lambda_2$ are the corresponding wavelengths, $c_{\rm int}$ is the intrinsic color ($J-H=-0.12$ or $H-K=-0.05$) of a B0 star from \citet{ko83}.

\begin{figure}
\includegraphics[width=0.5\textwidth]{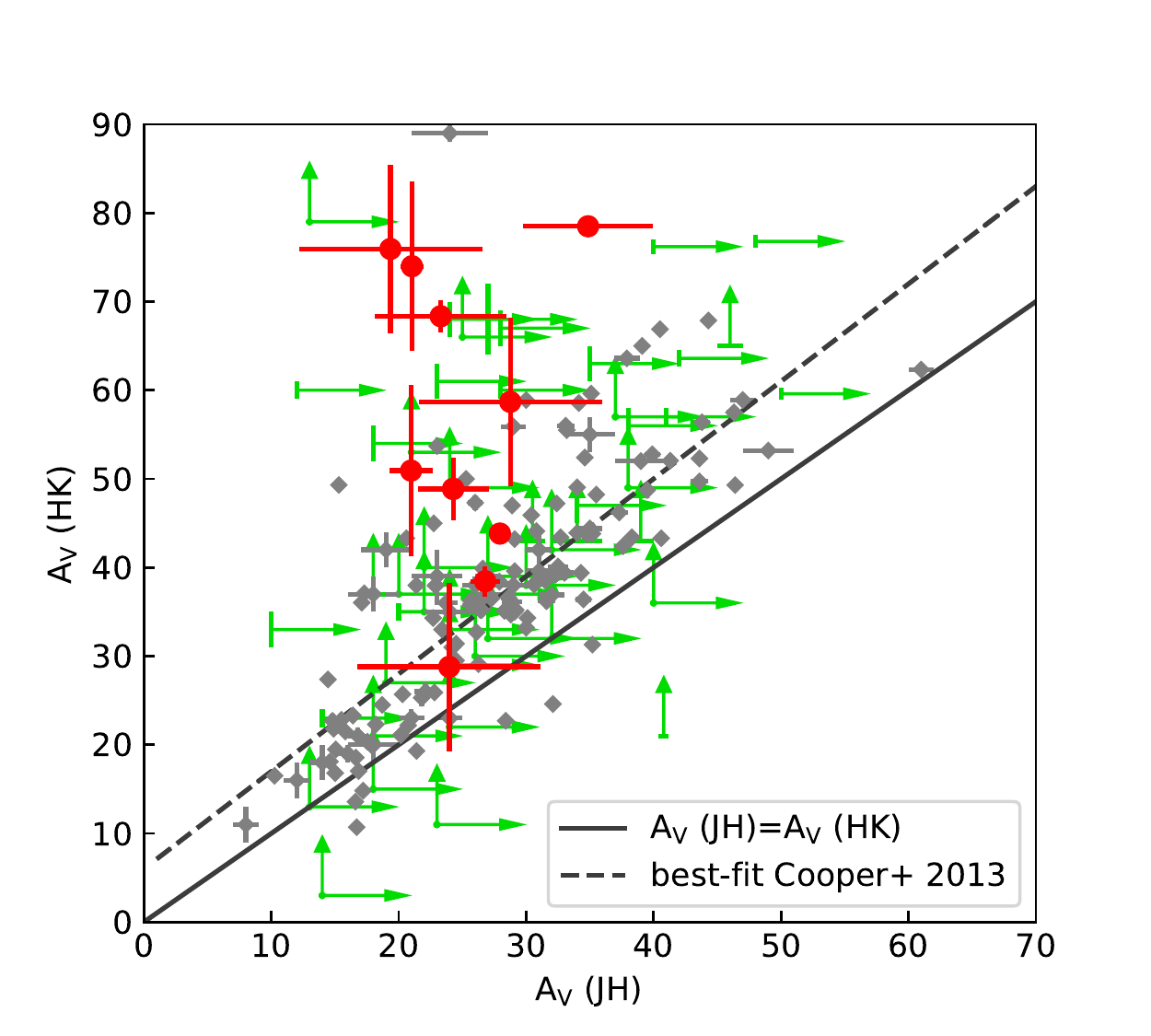}
\caption{Correlation of extinction derived using $JH$ and $HK$ colors. As for Figure \ref{fig:ccd}, the data points for our sources (see Table \ref{tab:2mass} for $JHK$ magnitudes) are shown in red and those in green/gray are for HM stars identified by \citet{co13} based on the RMS survey and follow-up near-IR spectroscopy. The extinction derived using two different pairs of bands are equal at the black solid line. The dashed line shows the linear regression line of the Cooper data derived by the authors \citep[Akritas–Thiel–Sen regressions; see][for details]{co13}.
}
\label{fig:av}
\end{figure}

In Figure \ref{fig:av},
we plot $A_{ V} (JH)$ versus $A_{ V} (HK)$ of our sources 
together with those of the \citet{co13} sample.
The sources observed by \citet{co13} show approximately linear correlation between $A_{ V} (JH)$ and $A_{ V} (HK)$, but the latter is systematically higher by a factor of typically $\sim$1.5.
In contrast, our sources have $A_{ V} (HK)$ significantly larger than $A_{ V} (JH)$, up to a factor of $\sim$4, due to the $K$-band excess shown in Section \ref{sec:sample}.
Therefore, we will adopt $A_{ V} (JH)$ to correct extinction in Section 5.2. These values are tabulated in Table \ref{tab:br}. We provide more complete information, including the 2MASS $JHK$ magnitudes and $A_{ V} (HK)$, in Appendix \ref{sec:appendix:2mass}.

One might be concerned that we cannot apply the above conventional extinction law, in particular if we observe the above fluxes via scattering on the disk or in an outflow cavity. Actually, for some HPMOs well-studied using millimeter interferometry, we do not seem to be able to directly observe light from the protostars in the infrared \citep[e.g.,][]{ta12}. In this case, the target emission must be extended. However, only a few of our targets were marginally resolved during our observations: the remaining targets are not resolved at the given seeing (0\farcs6-0\farcs8, corresponding to 700-3700 au). Observations at a higher angular resolutions (e.g., by using James Web Space Telescope) would allow us to better investigate this issue.


\subsection{Br $\gamma$ and High-Mass Star Formation}
\label{sec:im_br}

As shown in Section \ref{sec:result:Br}, the observed Br $\gamma$ emission shows a  triangular profile with a FWHM of 70--400 km s$^{-1}$, with most cases ranging from $\sim100-200~{\rm km~s^{-1}}$.
These are similar to the Br $\gamma$ profiles observed toward low-mass PMS stars \citep{fo01}.
For low-mass PMS stars, these line luminosities are well correlated with the disk accretion rates \citep{mu98}, and the emission is highly likely to be associated with magnetospheric accretion columns \citep{ca00}.

One might think that the Br $\gamma$ emission
for intermediate protostars to HMPOs could have a different origin.
\citet{ta16} presented modeled profiles of Br $\gamma$ emission
for Herbig AeBe stars associated with the following:
(1) a magneto-centrifugal disk wind, (2) an accretion disk heated by the illumination of the central source (hereafter ``a passive disk"), (3) a disk heated by accretion viscosity (hereafter ``an active disk"), and (4) magnetospheric accretion. 
However, these line profiles cannot explain the observed profiles adequately, as described below.
For three of the above four cases, a symmetric line profile with a single peak would require a very specific viewing angle: $i\ll20$\arcdeg~(nearly pole-on) for the passive disks and the disk winds; and $\sim 60$\arcdeg~for magnetospheric accretion.
These trends do not match the the following observations: among the sources with Br-$\gamma$ emission, disk inclination angles of 30\arcdeg-60\arcdeg~were measured for three of them (AFGL 490, W33A-MM1 Main and IRAS 18151-1208) using millimeter interferometry (Table \ref{tab:tar}).

The active disk models would, on the other hand, yield a symmetric line profile with a single peak over a large range of viewing angles ($i < 40$\arcdeg). However, their FWHMs are up to $\sim$50 km s$^{-1}$, significantly smaller than the observations. Furthermore, these models would yield a triangular profile over a very small range of inclination angles ($i\ll20$\arcdeg): the remaining line profiles resemble a Gaussian profile with blueshifted and redshifted wing components. This, again, 
is not consistent with the observations of the disk inclination angles described above.

Based on the line profiles similar to those of  low-mass PMS stars,
we assume that the Br $\gamma$ emission
is associated with magnetospheric accretion. Then,
we estimate the accretion luminosities and disk accretion rates
as follows.
We first derive the Br$\gamma$ luminosity using the observed equivalent width and the 2MASS $K$ magnitude, correcting with extinction derived using the $JH$ magnitudes (i.e., $A_{\rm v}$ (JH); see Section \ref{sec:av}).
The accretion luminosity is then estimated
using the same method as \citet{po17}. We first use the empirical relationship between $L_{\rm Br\gamma}$ and the accretion luminosity $L_{\rm acc}$ for Herbig Ae and Be stars \citep{me11}:
\begin{equation}
\log(\frac{L_{\rm acc}}{L_\odot})=(3.55\pm0.80)+(0.91\pm0.27)\log(\frac{L_{\rm Br\gamma}}{L_\odot})
\end{equation}
Then the disk accretion rate $\dot{M}_{\rm acc}$ is obtained using the following equation:
\begin{equation}
\dot{M}_{\rm acc}=\frac{L_{\rm acc}R_\star}{GM_\star}
\end{equation}
where $R_\star$ is the stellar radius, $M_\star$ is the stellar mass, and $G$ is the gravitational constant.
Here we adopt the stellar mass from the models of the zero-age-main-sequence-star in \citet{da11} given the $L_{\rm bol}$ or observational estimates (Table \ref{tab:tar}).
The stellar radius $R_\star$ is then obtained from the same models. It is noteworthy that the stellar radius might increase by a factor of a few to $\sim100$ for a massive protostar at an accretion phase \citep{ho10}. If this is the case, the mass accretion rate is underestimated, which might explain the mass accretion rate problem (see below).

Figure \ref{fig:macc} shows that the disk accretion rate in comparison with a possible evolutionary scheme:
i.e., No \ion{H}{2} region $\rightarrow$ HC \ion{H}{2} $\rightarrow$ UC \ion{H}{2} \citep{be05}.
No clear correlation is found in between the mass accretion rate and evolution.
The derived disk accretion rates ranges between $\lesssim10^{-8}$ and $\sim10^{-4}$ $M_\odot$ yr$^{-1}$.
Most of these accretion rates are not sufficient for forming
massive O stars ($M_* \gtrsim 10~M_\sun$)
in their formation timescale of $\sim10^{5}$ yr \citep{mc03}.
A possible explanation is that these 
stars have already accreted most of their mass.
If this is the case, our selected sources, as bright in infrared, have completed their main-accretion phase. This suggests that the aforementioned evolutionary scheme is not plausible for all HMPOs.
Alternatively, 
HM stars are associated with accretion bursts, which may be responsible for low-mass protostars accreting a significant fraction of their final mass
\citep{au14,hs18,hs19}.
In fact, a variation of mass accretion implied by a luminosity outburst was observed toward the HM protostar S255IR-SMA1 \citep{ca16,li18}.

\citet{li20} estimated an envelope mass infalling rate of $\sim10^{-4}~M_\odot~{\rm yr}^{-1}$ at $r$=200-100 au for S255IR SMA1. This is significantly larger than the upper limit of the disk accretion rate of $\sim$10$^{-8}$ $M_\odot~{\rm yr}^{-1}$ we measured using the Br-$\gamma$ emission. Such discrepancies between the envelope and disk accretion rates have been recognized for low-mass protostellar evolution, and support an evolutionary scenario with accretion outbursts triggered by an enhanced disk mass \citep{ca00}.

\begin{figure}
\includegraphics[width=0.45\textwidth]{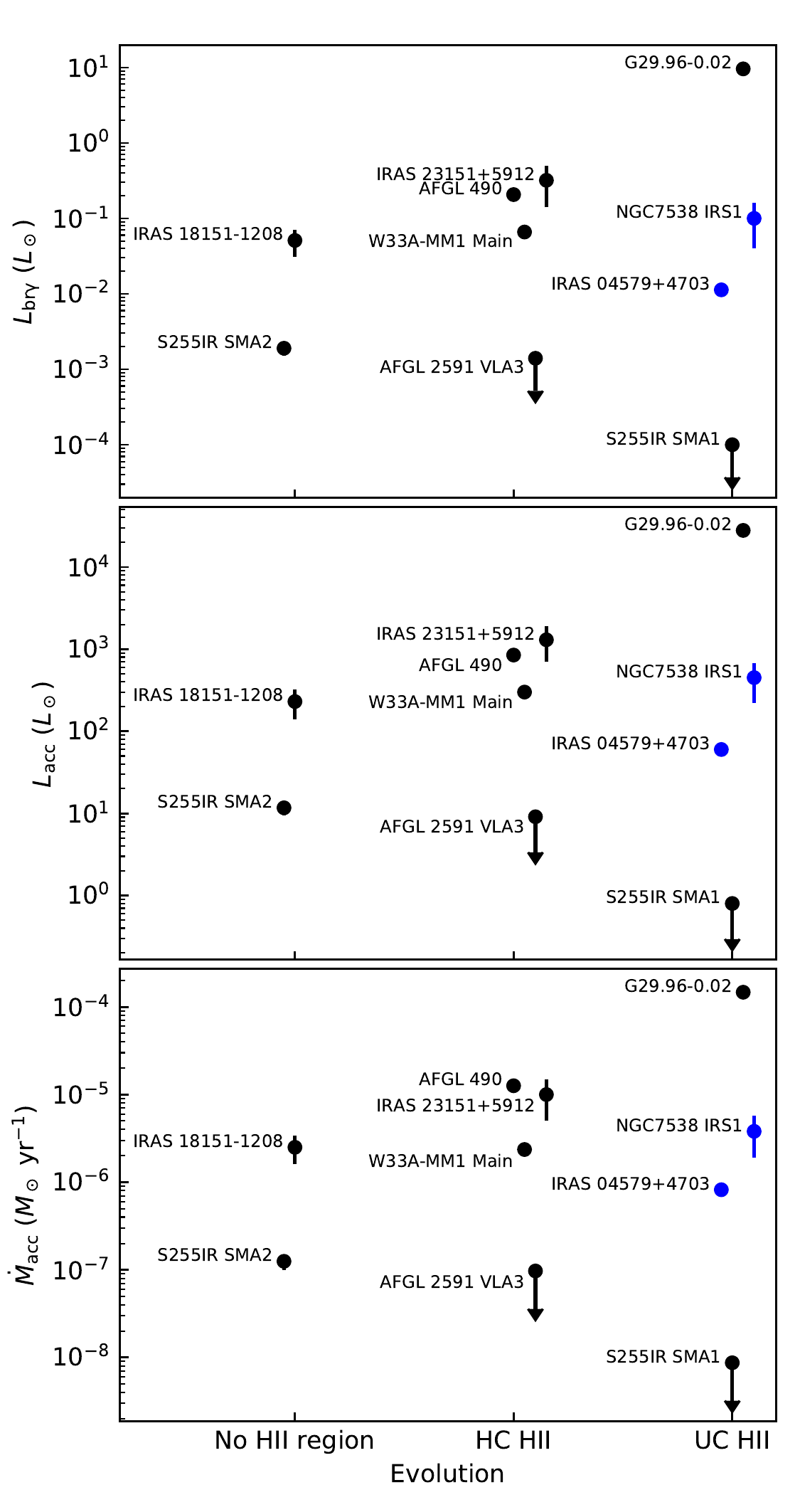}
\caption{Br$\gamma$ luminosity ({\it top}), accretion luminosity ({\it middle}), and disk accretion rate ({\it bottom}) versus associated \ion{H}{2} regions. The blue points are for sources associated with a possible HC/UC \ion{H}{2} region (see Table \ref{tab:tar}).
}
\label{fig:macc}
\end{figure}

As shown in Section \ref{sec:result:Br}, the Br $\gamma$ emission toward G29.96-0.02, the most massive source in our target list, exhibits a large equivalent width ($\sim-89~$\AA) and a narrow linewidth ($28~{\rm km~s^{-1}}$).
\citet{ma03} found extended Br $\gamma$ emission and suggested that such emission is caused by an ionizing star with a spectral type of O5$-$O6 V.
Of our sources, G29.96-0.02 is the only source with a detection in the Hydrogen Pfund series (Figure \ref{fig:co}) and is the most massive among sources associated with an UC \ion{H}{2} region (Table \ref{tab:tar}).
These peculiarities in the observed spectra suggest that this is the unique source for which the Br $\gamma$ emission is associated with an extended \ion{H}{2} region rather than a compact ionizing region in accretion flows.

As shown in Figure \ref{fig:br},
Br$\gamma$ absorption is detected towards only W3 IRS 5 among our sources (Figure \ref{fig:br}), which was not identified by the previous median-resolution observation ($R=4000$) in \citet{bi12}.
No clear emission feature was found 
by either \citet{bi12} or this work, implying that the accretion has stopped (or temporarily halted).
A Br $\gamma$ absorption with a narrow linewidth were found to originate from the photospheric absorption in main-sequence stars \citep{ha96,wa97}. However, it is found that the linewidth of the absorption usually exceeds 100 km s$^{-1}$ toward massive (O-type) main sequence stars \citep{ha05}, much larger than that in W3 IRS5 ($\Delta V\sim37~{\rm km~s^{-1}}$). Therefore, the origin of the absorption in W3 IRS5 is still unclear.

\subsection{CO overtone emission}

There is growing consensus that, for low-mass YSOs, the CO bandhead emission at 2.3--2.4 \micron~is associated with the the inner disk region where dust grains are sublimated \citep{du10}. High-resolution spectroscopy of the band emission profiles for some objects show double-peak line profiles and/or a blue-shoulder that are well explained by the disk models \citep[see][for review]{na07}. Such studies have also been extended toward young stars with intermediate-to-high masses \citep{il13,il14,mu13}.
Medium-to-high resolution spectra of some HMPOs also show band emission with a distinct profile indicating disk rotation \citep{il13,mu13}.

In both low-mass and high-mass cases, the observed CO fluxes or luminosities 
appear to be 
correlated with those of Br $\gamma$ when both emission features are detected
\citep{co10,il14,po17}.
However, the CO bandhead emission usually has a detection rate significantly lower than the Br $\gamma$ emission:
$\sim 20~\%$ (CO) and $\sim 80~\%$ (Br $\gamma$) for observations of low-mass Class I protostars by \citet{co10}, 
$7~\%$ and $70~\%$ for observations of Herbig AeBe stars by \citet{il14}, and
17-34 \% and 74-97 \% for observations of massive protostars by \citet{co13}, and \citet{po17}.

In our observations of eleven massive YSOs, the CO bandhead emission $v=2-0$ was detected only in W33A-MM1 Main and was marginally seen in IRAS 18151-1208, 
yielding a detection rate of only $9-18\%$.
A detection rate lower than \citet{co13,po17} is 
probably due to the lower signal-to-noise of our observations, which are
partially or fully attributed to 
the following:
(1) a spectral resolution significantly higher than previous observations \citep[$R$=35000-70000, 500 and 7000 for this work,][respectively]{co13,po17}; and
(2) a relatively large continuum excess emission (Section \ref{sec:av}).
For the former, the emission lines are significantly blended at the bandhead, making the feature broad \citep[$>0.002~ \micron$; e.g.,][; see also the spectrum of W33A-MM1 Main in Figure 3]{il13}, and a very high angular resolution splits this emission into many spectral pixels, yielding a low signal-to-noise.
We smoothed the spectra to increase signal-to-noise, but did not find clear detection towards more sources.

We measure an extinction-corrected CO luminosity of W33A-MM1 Main of 0.04$\pm$0.01 $L_\odot$.
The CO/Br-$\gamma$ luminosity ratio of $\sim$0.6 roughly agrees with the empirical correlation for Herbig AeBe stars and massive protostars shown in \citet{po17}.

\citet{il18} used LTE analytic disk models to explain the low detection rates of the CO overtone emission. These authors demonstrated
that the CO overtone emission would be observed only over a specific range of disk accretion rates. The emission is considered to be associated with the gaseous inner disk where dust is sublimated. A low disk accretion rate ($\dot{M} \lesssim 10^{-6}~M_\sun$ yr$^{-1}$) would result in a small dust-sublimation region in the inner disk, a correspondingly smaller emitting area for CO, and therefore fainter CO emission with respect to the continuum.
On the other hand, a high disk accretion rate ($\dot{M} \gtrsim 10^{-4}~M_\sun$ yr$^{-1}$) would yield not only a large dust-sublimation region in the inner disk (and therefore the CO emitting region), but also intense thermal dust continuum emission, making the CO emission lower with respect to the continuum.

As derived in Section \ref{sec:im_br}, the disk accretion rates inferred from the Br-$\gamma$ emission range from $<$10$^{-8}$ to $\sim$10$^{-4}$ $M_\sun$ yr$^{-1}$ (Table \ref{tab:br}, Figure \ref{fig:macc}). Among these objects, intermediate disk accretion rates (2-3$\times$10$^{-6}$ $M_\sun$ yr$^{-1}$) were measured for W33A-MM1 Main and are marginally seen in IRAS 18151-1208, i.e., those with detection of the CO bandhead emission. In this context, our detection would be broadly consistent with the model predictions by \citet{il18}; the CO bandhead EWs are at maximam values for $\dot{M}_{\rm acc}\sim10^{-5}~M_\odot~{\rm yr}^{-1}$. However, a few objects have similar disk accretion rates (e.g., 2-6 $\times$10$^{-6}$ $M_\sun$ yr$^{-1}$ for NGC 7538 IRS1, Table \ref{tab:br}) but without CO detection. We speculate that their negative detection can be attributed to a disk inclination angle close to edge-on, making the CO emission region undetectable. Even so, we detected CO emission in only two of seven objects with disk accretion rates between $\sim$10$^{-6}$ and $\sim$10$^{-4}$ $M_\sun$ yr$^{-1}$, for which \citet{il18} predicted relatively large band-to-continuum ratios. A large sample with better signal-to-noise ratio is required to further address this issue.


\subsection{Nature of CO $v$=0-2 Absorption}
\label{sec:origin}

As shown in Section \ref{sec:co abs}, the CO absorption observed toward many of our sources is either blueshifted or redshifted, suggesting that the absorption is associated with a relatively cool outflow or inflow. This is in contrast to the discussion in previous studies of a few HM stars, for which the authors attributed the absorption to a circumatellar disk.
\citet{da10,mu13} observed CO absorption toward a few protostars using integral field spectroscopy with a modest spectral resolution ($R\sim$ 5500, yielding a velocity resolution of $\sim 55$ km s$^{-1}$). These authors analyzed the spatial distribution of the velocity centroids of the absorption lines, and revealed kinematics similar to disk rotation.
Assuming that the observed kinematics of $\Delta v = $5-10 km s$^{-1}$ is due to Keplerian rotation, they derived central masses of these objects of 15--30 $M_\sun$. 

In Section \ref{sec:co abs} we measured the absorption equivalent widths of the individual transitions associated with most of our target sources. We apply a simple model to derive the temperature and column density, and discuss the origin of this cool gas component.
For a given transition, the absorption line profile $I_\nu$ is expressed as
\begin{equation}
I_\nu=I_{\rm c}\exp(-s \phi_\nu N_{v, J}),
\label{eq:abs}
\end{equation}
where $I_{\rm c}$ is the continuum flux, $s$ is the integrated cross section, $\phi_\nu$ is the normalized Gaussian function and $N_{v, J}$ is the column density in the lower energy state \citep{sm09}. Under optically thin conditions, the equation can be re-written as 
\begin{equation}
\frac{EW}{\lambda}=s N_{v, J},
\end{equation}
where $EW$ and $\lambda$ are the equivalent width and wavelength, respectively 
\citep{go03}.

\begin{figure}
\includegraphics[width=0.5\textwidth]{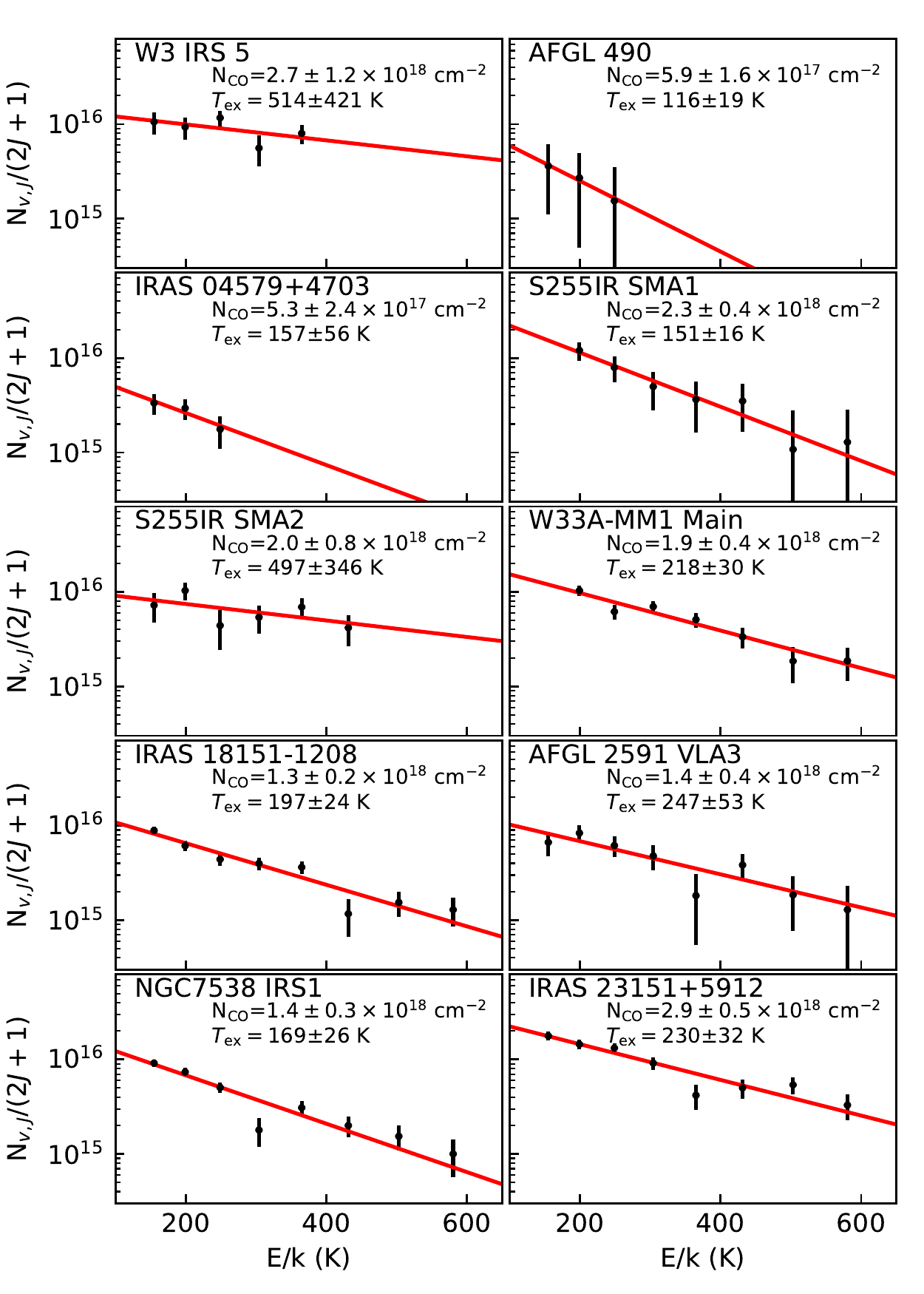}
\caption{Boltzmann plots of the CO ro-vibrational line absorption.
The red lines show the best-fit
LTE models with the total CO column density and the temperature described in individual panels.
}
\label{fig:bol}
\end{figure}

We derived $N_{v, J}$ using Equation (5) for individual sources and transitions, and plot them in Figure \ref{fig:bol}.
Using scipy.optmize.curve\_fit, we fit these level populations
assuming that the CO absorption is associated with
an optically-thin LTE gas layer with a single temperature ($T_{\rm}$).
These indicate a temperature for the CO absorption region of 100-200 K for all the sources but W3 IRS 5 and S255IR SMA2, for which the derived temperatures have large uncertainties; these large uncertainties come from the relatively high noise level of the measured EWs (Table \ref{tab:co}). 

Our successful fitting with an LTE model indicates
that the gas components observed at these transitions are well thermalized with a density higher than the critical density $n_{\rm H_2}\sim5\times10^{5}~{\rm cm^{-3}}$ of CO at $v=0$, $J=8-15$ \citep{ya10}.
Given CO column densities of $\sim(0.5-2.9)\times10^{18}~{\rm cm^{-2}}$, the thicknesses of the warm gas components are less than
$700-3900$ au taking the canonical CO/$H_2$ abundance of $\sim10^{-4}$.
If the warm gas component has a higher density, e.g., $\sim10^7~{\rm cm^{-3}}$, the absorption would trace an inner region within $30-200$ au.

Given the temperature of these gas components, we estimate their distance to the central heating source. Given the source luminosity (Table \ref{tab:tar}), the dust temperature can be expressed as a function of radius if we assume that the power radiated is equal to that absorbed in a spherical symmetric cloud (i.e., $T_{\rm d}\propto \frac{L^{1/5}}{R^{2/5}}$, see equation 31 in \citet{go74}). As a result, we estimate that the absorption occurred inside a region with a typical radius of a few hundred aus ($\lesssim200-600$ au. depending on source), except for W3 IRS 5; with the large uncertainty in temperature, the radius of the CO component is not constrained well at $\sim10-3000$ au.

In practice, absorption components at different velocities would be associated with gas at different radii. Figure \ref{fig:stack} shows that S255 SMA1 and SMA2 are associated with redshifted CO absorption up to $\sim$30 km s$^{-1}$. One would estimate that the absorption component at the highest velocity would be associated with $r \sim 6$ au, assuming free fall motion towards the HPMO with a stellar mass of 20$M_\sun$. Observations at a higher signal-to-noise would allow us to better investigate temperature distributions of CO gas at different radii .


\section{Summary}
\label{sec:summary}

We have obtained high-resolution $K$-band spectra ($R$=35,000 or 70,000) for eleven HM (proto-)stars associated with a disk or a disk candidate observed at submillimeter-to-radio wavelengths. Their bolometric luminosities and associated \ion{H}{2} region indicate that these sources have a variety of stellar masses ($M_\star=7-56~M_\odot$) and evolutionary stages.
In this paper, we focus our analysis and discussion on the CO overtone and Br $\gamma$, whose emission potentially trace mass accretion in the inner disk region (r$\ll$100 au) and the inner disk edge to the star, respectively.

Br-$\gamma$ emission was observed toward eight sources. Most of them show a line width of typically 100-200 km s$^{-1}$, and their triangular line profiles are similar to those of low-mass PMS stars, for which the Br-$\gamma$ emission is associated with magnetospheric accretion from the inner disk edge to the star. 
As for previous studies, we extended a correlation between the Br-$\gamma$ and accretion luminosities for PMS stars and estimated the disk accretion rates of our sources. These range from $\lesssim$10$^{-8}$ and $\sim$10$^{-4}$ $M_\sun$ yr$^{-1}$, 
without any correlation with a possible evolutionary sequence based on an associated \ion{H}{2} region (no \ion{H}{2}$\rightarrow$HC \ion{H}{2}$\rightarrow$UC \ion{H}{2}). The derived disk accretion rates are too small for each HM star to reach its final mass within the predicted formation time scale of $\sim 10^5$ yrs. This suggests that these sources have already gained most of the final stellar masses, or gain more mass via episodic accretion as proposed for low-mass protostars. 
For one of the sources, the inferred disk accretion rate contrasts to a large envelope accretion rate measured at millimeter wavelengths, as for some low-mass protostars, supporting the latter scenario.

The Br-$\gamma$ emission associated with G29.96-0.02, which is characterized by a narrow line width ($V_{\mathrm FWHM} \sim 30$ km s$^{-1}$) is attributed to an extended \ion{H}{2} region ($\sim0.2$ pc, \citealt{ma03}). W3 IRS 5 alternatively exhibits Br-$\gamma$ absorption presumably associated with the stellar photosphere.

CO overtone emission was observed only toward two sources, with one a marginal detection. A detection rate significantly lower than for Br $\gamma$ is consistent with previous observations of CO overtone emission toward a number of HM (proto-)stars, and the theory predicting that the band-to-continuum ratio is high only for a specific range of disk accretion rates.

All sources except G29.96-0.02 exhibit CO $v$=2-0 absorption. Most of them are either blueshifted or redshifted, indicating that the absorption is associated with an outflow or an inflow with a radial velocity of $\sim5-30$ km s$^{-1}$. Their Boltzmann plots indicate that gas associated with CO $v$=2-0 absorption is in LTE conditions (and therefore $n_{\mathrm H_2} \gtrsim 10^6$ cm$^{-3}$) at a single temperature of typically 100-200 K. Together with the luminosity of individual sources, we derive a typical upper-limit distance from the star to the CO absorption layer of 200-600 au.


\acknowledgments
We express our gratitude to the anonymous referee for the constructive comments that improved this paper.
We thanks
We thanks the staff at the Infrared Telescope Facility (IRTF) which is operated by the University of Hawaii under contract 80HQTR19D0030 with the National Aeronautics and Space Administration.
This research has been supported by the Ministry of Science and Technology (MoST) of Taiwan (grant No. 106-2119-M-001-026-MY3).
M.T. is supported by JSPS KAKENHI grant Nos. 18H05442, 15H02063, and 22000005.


\appendix
\setcounter{figure}{0}
\setcounter{table}{0}
\renewcommand{\thefigure}{A\arabic{figure}}
\renewcommand{\thetable}{A\arabic{table}}
\renewcommand{\thesection}{\Alph{section}}


\begin{deluxetable*}{cccccc}
\tabletypesize{\footnotesize}
\tablecaption{2MASS magnitudes and extinction values}
\tablehead{ 	
\colhead{Name}
& \colhead{$J$}		
& \colhead{$H$}
& \colhead{$K$}	
& \colhead{A$_{\rm V}$ ($JH$)}	
& \colhead{A$_{\rm V}$ ($HK$)}
\\
\colhead{}		
& \colhead{[mag]}
& \colhead{[mag]}	
& \colhead{[mag]}		
& \colhead{[mag]}
& \colhead{[mag]}
}
\startdata 
W3 IRS 5        & 17.95$^*$     & 15.93$^*$     & 11.86$\pm$0.04        & 19.35$\pm$7.19        & 75.94$\pm$9.50        \\
AFGL 490        & 10.95$\pm$0.02        & 8.08$\pm$0.04 & 5.72$\pm$0.02 & 27.94$\pm$0.46        & 43.83$\pm$0.88        \\
IRAS 04579+4703 & 15.69$\pm$0.12        & 13.51$\pm$0.11        & 10.77$^*$     & 20.97$\pm$1.68        & 50.93$\pm$9.69        \\
S255IR SMA1     & 15.46$\pm$0.22        & 12.95$\pm$0.16        & 10.32$\pm$0.09        & 24.27$\pm$2.78        & 48.87$\pm$3.56        \\
S255IR SMA2     & 15.46$\pm$0.22        & 12.95$\pm$0.16        & 10.32$\pm$0.09        & 24.27$\pm$2.78        & 48.87$\pm$3.56        \\
W33A-MM1 Main   & 15.34$\pm$0.06        & 13.15$\pm$0.06        & 9.20$^*$      & 21.04$\pm$0.91        & 73.99$\pm$9.54        \\
IRAS 18151-1208 & 15.32$^*$     & 12.91$\pm$0.09        & 9.25$\pm$0.03 & 23.29$\pm$5.16        & 68.35$\pm$1.80        \\
G29.96-0.02     & 13.80$\pm$0.09        & 11.05$\pm$0.07        & 8.97$\pm$0.06 & 26.78$\pm$1.16        & 38.45$\pm$1.72        \\
AFGL 2591 VLA3  & 14.32$^*$     & 10.77$\pm$0.04        & 6.58$\pm$0.02 & 34.86$\pm$5.10        & 78.50$\pm$0.85        \\
NGC 7538 IRS1    & 14.59$^*$     & 11.64$^*$     & 8.49$\pm$0.04 & 28.75$\pm$7.19        & 58.69$\pm$9.50        \\
IRAS 23151+5912 & 12.78$^*$     & 10.30$^*$     & 8.73$\pm$0.05 & 23.96$\pm$7.19        & 28.80$\pm$9.51
\enddata
\tablecomments{$^*$ a magnitude without an uncertainty in 2MASS catalog. A large uncertainty of 0.5 is used to calculate $A_{\rm V}$.
}
\label{tab:2mass}
\end{deluxetable*} 

\section{2MASS Magnitudes and Extinction}
\label{sec:appendix:2mass}
In Section \ref{sec:av} we estimated extinction based on the 2MASS $JHK$ magnitudes.
Table \ref{tab:2mass} shows the $JHK$ magnitudes of the individual sources and extinction derived using $JH$ and $HK$ magnitudes. See Section \ref{sec:av} for other details. 
The uncertainties are not given in the 2MASS catalog for some magnitudes for which we conservatively assume a large uncertainty (0.5 magnitude) in our analysis.
Note that S255IR SMA1/SMA2 share the same color (as well as the extinction, section \ref{sec:av}) as they were not resolved in the 2MASS images.


\section{Comparisons with previous B\lowercase{r}$\gamma$ observations}
\label{app:br}
\subsection{AFGL 490/IRAS 18151-1208/AFGL 2591 VLA3/NGC 7538 IRS1/IRAS 23151+5912}
The equivalent widths of Br $\gamma$ line emission have been measured by the surveys of \citet{co13} and/or \citet{po17}.
The measured equivalent widths are -4.8, -1.4, -0.3, -5.1 and -4.0 \AA~for AFGL 490, IRAS 18151-1208, AFGL 2591 VLA3, NGC 7538 IRS1, and IRAS 23151+5912, respectively.
These measurements are consistent with ours within a factor of $\sim$2.5 except for AFGL 2591 VLA3.
While \citet{po17} found an equivalent width of $-0.3$ \AA~toward AFGL 2591 VLA3, we give an upper limit of $0.03$ \AA~(Figure \ref{fig:br}).

\subsection{W3 IRS 5}
\citet{bi12} conducted a $K_s$-Band spectroscopic survey of young stars and YSOs in the W3-Main star forming regions in 2009. These authors did not detect Br-$\gamma$ emission associated with W3 IRS 5.
In our high-resolution spectrum, we found a Br $\gamma$ absorption feature at the systemic velocity with an equivalent width of $0.58\pm0.04$ \AA~and a narrow linewidth of $36.6~{\rm km~s^{-1}}$.

\subsection{IRAS 04579+4703}
\citet{is01}
 measured the equivalent width of Br$\gamma$ to be $-12.6$ \AA~which is larger than our derived value by a factor of $\sim$2.3 (Table \ref{tab:br}).

\subsection{S255IR SMA1/SMA2}
As mentioned in \ref{sec:s255} S255IR SMA1 experienced a recent accretion burst between 2007 and 2016 \citep{ca16,li18}.
Due to high extinction and/or strong veiling, no clear Br $\gamma$ emission was detected toward the source center (S255IR IRS3 in \citealt{ca16}), consistent with our result in S255IR SMA1.

\subsection{W33A-MM1 Main}
The equivalent width of Br$\gamma$ is measured to be $-5$ \AA~and $-3.17$ \AA~in \citet{da10} and this work, respectively.
This small difference might come from the different spectral resolutions or time variability.

\subsection{G29.96-0.02}
G29.96-0.02 shows distinct features from our survey sample. 
It has an unusual strong Br $\gamma$ line emission with an equilvalent width of $\sim-$90 \AA~(Table \ref{tab:br}).
Such strong Br $\gamma$ emission was also observed in \citet{ma03} with the median-resolution observation using VLT.
In addition, several helium recombination lines were detected in the literature.
Pfund series emission is also detected in G29.96-0.02 (Figure \ref{fig:co}).
These results suggest that G29.96-0.02 contains a large amount of highly excited material.
The peculiarity of this source is probably related to the fact that it is the brightest ($L=8\times10^5 L_\odot$) and the most massive ($M_{\rm star}\sim33-56 M_\odot$) of our sources (Table \ref{tab:tar}).


\section{Comparisons with previous observations of the CO \lowercase{$v=2-0$} bandhead emission}

\subsection{AFGL 490/IRAS 18151-1208/AFGL 2591 VLA3/NGC 7538 IRS1/IRAS 23151+5912}
Ks-band spectroscopic observations covering CO $v$=0-2 were taken by \citet{co13} and/or \citet{po17} in these sources, and thus we discuss them concurrently.
CO bandhead emission is not detected in these sources in the literature.
\citet{co13} and \citet{po17} executed spectroscopy for these sources, but they did not detect the CO bandhead emission.
These non-detections are mostly consistent with our work (except for IRAS 18151-1208, see \ref{sec:co18151}) as our observations provide a higher spectral resolution but a lower sensitivity.

\subsection{W3 IRS 5}
\citet{bi12} conducted $K$s-Band spectroscopic observations toward the sources in the W3-Main star forming region.
Among their targets, 15 candidates were identified as OB stars through spectral classification. 
The photospheres of OB stars are detected from the more evolved diffuse \ion{H}{2} regions, indicating that most of these OB stars have finished their star formation. Among their targets, W3 IRS 5 is the only source that is found to be still accreting. CO bandhead emission was not detected in \citet{bi12} or in our work (with an rms noise level of 3 \AA~integrating over $2.289-2.313~\micron$).


\subsection{IRAS 04579+4703}
\citet{is01} did not detect CO bandhead emission.
While \citet{is01} measured the upper limit of the CO $v=2-0$ bandhead emission to be $>-6.2$ \AA, our data gives a constraint of $>-6.4$ \AA~given the integrated range of $2.289-2.313~\mu$m.

\subsection{S255IR SMA1/SMA2}
\label{sec:s255}
S255IR SMA1 has been found to have experienced an accretion burst on the basis of near-infrared spectroscopic observations in 2007 and 2016 \citep{ca16}.
Soon after that, \citet{li18} reported that the submillimeter continuum emission dimmed from 2016 to 2017.
\citet{ca16} show the spectra of the source and the outflow cavity in the pre-burst phase (2007) and the outburst phase (2016).
As a result, the CO bandhead emission is only seen towards the outflow cavity in the pre-burst phase.
This is consistent with our result of non-detection toward the source as also reported in \citet{co13} from observation in 2005.

\subsection{W33A-MM1 Main}
CO bandhead emission is detected in \citet{da10} and \citet{il13} as well as in our observations.
With high-resolution observation ($R\sim30,000$), \citet{il13} find that CO bandhead emission is well fitted by a model of a Keplerian rotation disk;
the CO emission traces the disk from the inner region from a few au to a few tens of au.
In our spectra we measured an equivalent width of the CO $v$=2--0 bandhead emission of $-2.33\pm0.69$ \AA.

\subsection{IRAS 18151-1208}
\label{sec:co18151}
The $v=2-0$ and $3-1$ CO bandhead emission was observed by \citet{mu13}.
At a lower spectral resolution 
($R$$\sim$5500),
\citet{da04} detected strong CO bandhead emission between $v=2-0$ and $v=7-5$.
This observations were taken with a much lower resolution and the emission is interpreted as inner disk surface or outer accreting funnel-flow.
In \citet{co13}, the CO emission is not detected, likely due to the sensitivity. 

As shown in Section \ref{sec:result:coemission}, we have marginally detected the $v=2-0$ bandhead emission.
It is difficult to discuss detection of the CO $v=3-1$ bandhead emission since the continuum is not well defined (see Table \ref{fig:co}).


\bibliographystyle{aasjournal}

\begin{thebibliography}{}
\bibitem[Audard et al.(2014)]{au14} Audard, M., {\'A}brah{\'a}m, P., Dunham, M.~M., et al.\ 2014, Protostars and Planets VI, 387
\bibitem[Beck et al.(2010)]{be10} Beck, T.~L., Bary, J.~S., \& McGregor, P.~J.\ 2010, \apj, 722, 1360 
%
\bibitem[Beltr{\'a}n et al.(2011)]{be11} Beltr{\'a}n, M.~T., Cesaroni, R., Neri, R., Codella, C.\ 2011, \aap, 525, A151 
%
\bibitem[Beuther et al.(2012)]{be12} Beuther, H., Linz, H., \& Henning, T.\ 2012, \aap, 543, A88. doi:10.1051/0004-6361/201219128
\bibitem[Beltr{\'a}n et al.(2014)]{be14} Beltr{\'a}n, M.~T., S{\'a}nchez-Monge, {\'A}., Cesaroni, R., et al.\ 2014, \aap, 571, A52 
\bibitem[Beltr{\'a}n \& de Wit (2016)]{be16} Beltr{\'a}n, M.~T., \& de Wit, W.~J.\ 2016, \aapr, 24, 6
\bibitem[Beuther, \& Shepherd(2005)]{be05} Beuther, H., \& Shepherd, D.\ 2005, Astrophysics and Space Science Library, 105
\bibitem[Beuther et al. (2007)]{be07}Beuther, H., Zhang, Q., Bergin, E. A., et al. 2007, \aap, 468, 1045
\bibitem[Bik et al.(2005)]{bi05} Bik, A., Kaper, L., Hanson, M.~M., et al.\ 2005, \aap, 440, 121
\bibitem[Bik et al. (2012)]{bi12}Bik, A., Henning, T., Stolte, A., et al. 2012, \apj, 744, 87
\bibitem[Chabrier(2005)]{ch05} Chabrier, G.\ 2005, The Initial Mass Function 50 Years Later, 41
\bibitem[Chen et al. (2004)]{ch04}Xue-Peng Chen and Yong-Qiang Yao 2004 Chin. J. Astron. Astrophys. 4 284
\bibitem[Cooper et al.(2013)]{co13} Cooper, H.~D.~B., Lumsden, S.~L., Oudmaijer, R.~D., et al.\ 2013, \mnras, 430, 1125 
\bibitem[Calvet et al.(2000)]{ca00} Calvet, N., Hartmann, L., \& Strom, S.~E.\ 2000, Protostars and Planets IV, 377
\bibitem[Caratti o Garatti et al. (2016)]{ca16}Caratti o Garatti, A., Stecklum, B., Garcia Lopez, R., et al. 2016, NatPh, 13, 276
\bibitem[Claussen et al.(1994)]{cl94} Claussen, M.~J., Gaume, R.~A., Johnston, K.~J., et al.\ 1994, \apjl, 424, L41. doi:10.1086/187270
\bibitem[Connelley \& Greene(2010)]{co10} Connelley, M.~S., \& Greene, T.~P.\ 2010, \aj, 140, 1214
\bibitem[Cushing et al.(2004)]{cu04} Cushing, M.~C., Vacca, W.~D., \& Rayner, J.~T.\ 2004, \pasp, 116, 362. doi:10.1086/382907
\bibitem[Davis et al.(2004)]{da04} Davis, C.~J., Varricatt, W.~P., Todd, S.~P., \& Ramsay Howat, S.~K.\ 2004, \aap, 425, 981 
\bibitem[Davies et al. (2010)]{da10}Davies, B., Lumsden, L. S., Hoare, M. G., Oudmaijer, R. D., \& de Wit, W. 2010, \mnras, 402, 1504
\bibitem[Davies et al.(2011)]{da11} Davies, B., Hoare, M.~G., Lumsden, S.~L., et al.\ 2011, \mnras, 416, 972
\bibitem[de Wit et al.(2010)]{de10} de Wit, W.~J., Hoare, M.~G., Oudmaijer, R.~D., et al.\ 2010, \aap, 515, A45. doi:10.1051/0004-6361/200913209
\bibitem[Dullemond \& Monnier(2010)]{du10} Dullemond, C.~P., \& Monnier, J.~D.\ 2010, \araa, 48, 205
\bibitem[Dullemond et al. (2012)]{du12}Dullemond, C. P., Juhasz, A., Pohl, A., et al. 2012, Astrophysics Source Code Library, ascl:1202.015
\bibitem[Fallscheer et al.(2011)]{fa11} Fallscheer, C., Beuther, H., Sauter, J., et al.\ 2011, \apj, 729, 66
\bibitem[Folha \& Emerson(2001)]{fo01} Folha, D.~F.~M., \& Emerson, J.~P.\ 2001, \aap, 365, 90
\bibitem[Galv{\'a}n-Madrid et al.(2010)]{ga10} Galv{\'a}n-Madrid, R., Zhang, Q., Keto, E., et al.\ 2010, \apj, 725, 17
\bibitem[Goldreich \& Kwan(1974)]{go74} Goldreich, P. \& Kwan, J.\ 1974, \apj, 189, 441
\bibitem[Goto et al.(2003)]{go03} Goto, M., Usuda, T., Takato, N., et al.\ 2003, \apj, 598, 1038 
\bibitem[Hoare et al.(2005)]{ho05} 
Hoare M. G. et al., 2005, in Cesaroni R., Felli M., Churchwell E.,Walmsley M., eds, IAU Symp 227, Massive Star Birth: A Crossroads of Astrophysics. Cambridge Univ. Press, Cambridge , p. 370
\bibitem[Hsieh et al.(2016)]{hs16} Hsieh, T.-H., Lai, S.-P., Belloche, A., et al.\ 2016, \apj, 826, 68
\bibitem[Hsieh et al.(2018)]{hs18} Hsieh, T.-H., Murillo, N.~M., Belloche, A., et al.\ 2018, \apj, 854, 15
\bibitem[Hsieh et al.(2019)]{hs19} Hsieh, T.-H., Murillo, N.~M., Belloche, A., et al.\ 2019, \apj, 884, 149
\bibitem[Hunter (2007)]{matplotlib}Hunter, J. D. 2007, Computing in Science and Engineering, 9, 90
\bibitem[Ishii et al.(2001)]{is01} Ishii, M., Nagata, T., Sato, S., et al.\ 2001, \aj, 121, 3191 
\bibitem[Ilee et al.(2013)]{il13} Ilee, J.~D., Wheelwright, H.~E., Oudmaijer, R.~D., et al.\ 2013, \mnras, 429, 2960 
\bibitem[Ilee et al.(2014)]{il14} Ilee, J.~D., Fairlamb, J., Oudmaijer, R.~D., et al.\ 2014, \mnras, 445, 3723 
\bibitem[Ilee et al.(2016)]{il16} Ilee, J.~D., Cyganowski, C.~J., Nazari, P., et al.\ 2016, \mnras, 462, 4386 
\bibitem[Ilee et al.(2018)]{il18} Ilee, J.~D., Oudmaijer, R.~D., Wheelwright, H.~E., \& Pomohaci, R.\ 2018, \mnras, 477, 3360 
\bibitem[Hanson et al.(1996)]{ha96} Hanson, M.~M., Conti, P.~S., \& Rieke, M.~J.\ 1996, \apjs, 107, 281
\bibitem[Hanson et al.(2005)]{ha05} Hanson, M.~M., Kudritzki, R.-P., Kenworthy, M.~A., et al.\ 2005, \apjs, 161, 154. doi:10.1086/444363
\bibitem[Harsono et al. (2015)]{ha15}Harsono, D., Bruderer, S., \& van Dishoeck, E. F. 2015, \aap, 582, A41
\bibitem[Hosokawa et al.(2010)]{ho10} Hosokawa, T., Yorke, H.~W., \& Omukai, K.\ 2010, \apj, 721, 478. doi:10.1088/0004-637X/721/1/478
\bibitem[Izquierdo et al.(2018)]{iz18} Izquierdo, A.~F., Galv{\'a}n-Madrid, R., Maud, L.~T., et al.\ 2018, \mnras, 478, 2505. doi:10.1093/mnras/sty1096
\bibitem[Johnston et al.(2013)]{jo13} Johnston, K.~G., Shepherd, D.~S., Robitaille, T.~P., et al.\ 2013, \aap, 551, A43. doi:10.1051/0004-6361/201219657
\bibitem[Johnston et al.(2015)]{jo15} Johnston, K.~G., Robitaille, T.~P., Beuther, H., et al.\ 2015, \apjl, 813, L19 \bibitem[Kahn(1974)]{ka74} Kahn, F.~D.\ 1974, \aap, 37, 149 
\bibitem[Koornneef(1983)]{ko83} Koornneef, J.\ 1983, \aap, 500, 247
\bibitem[Krumholz et al.(2007)]{kr07} Krumholz, M.~R., Klein, R.~I., \& McKee, C.~F.\ 2007, \apj, 656, 959
\bibitem[Krumholz et al.(2009)]{kr09} Krumholz, M.~R., Klein, R.~I., McKee, C.~F., Offner, S.~S.~R., \& Cunningham, A.~J.\ 2009, Science, 323, 754
\bibitem[Kuiper et al.(2010)]{ku10} Kuiper, R., Klahr, H., Beuther, H., \& Henning, T.\ 2010, \apj, 722, 1556
\bibitem[Kuiper et al.(2011)]{ku11} Kuiper, R., Klahr, H., Beuther, H., \& Henning, T.\ 2011, \apj, 732, 20 
\bibitem[Leurini et al.(2009)]{le09}Leurini, S., Codella, C., Zapata, L. A., et al. 2009, \aap, 507, 1443
\bibitem[Liu et al.(2018)]{li18} Liu, S.-Y., Su, Y.-N., Zinchenko, I., Wang, K.-S., \& Wang, Y.\ 2018, \apjl, 863, L12 
\bibitem[Liu et al.(2020)]{li20} Liu, S.-Y., Su, Y.-N., Zinchenko, I., et al.\ 2020, arXiv:2010.09199
\bibitem[Lord (1992)]{lo92}Lord, S. D. 1992, A new software tool for computing Earth’s atmospheric transmission of near- and far-infrared radiation, Tech. rep.
\bibitem[Mart{\'{\i}}n-Hern{\'a}ndez et al.(2003)]{ma03} Mart{\'{\i}}n-Hern{\'a}ndez, N.~L., Bik, A., Kaper, L., Tielens, A.~G.~G.~M., \& Hanson, M.~M.\ 2003, \aap, 405, 175 
\bibitem[McKee \& Tan(2003)]{mc03} McKee, C.~F., \& Tan, J.~C.\ 2003, \apj, 585, 850
\bibitem[Mendigut{\'\i}a et al.(2011)]{me11} Mendigut{\'\i}a, I., Calvet, N., Montesinos, B., et al.\ 2011, \aap, 535, A99
\bibitem[Muzerolle et al.(1998)]{mu98} Muzerolle, J., Hartmann, L., \& Calvet, N.\ 1998, \aj, 116, 2965 
\bibitem[Mohanty et al.(2005)]{mo05} Mohanty, S., Jayawardhana, R., \& Basri, G.\ 2005, \apj, 626, 498
\bibitem[Moscadelli \& Goddi(2014)]{mo14} Moscadelli, L. \& Goddi, C.\ 2014, \aap, 566, A150
\bibitem[Murakawa et al. (2013)]{mu13}Murakawa, K., Lumsden, S. L., Oudmaijer, R. D., et al. 2013, \mnras, 436, 511
\bibitem[Najita et al.(1996)]{na96} Najita, J., Carr, J.~S., \& Tokunaga, A.~T.\ 1996, \apj, 456, 292 
\bibitem[Najita et al.(2000)]{na00} {Najita}, J.~R. and {Edwards}, S. and {Basri}, G. and {Carr}, J.\ 2000, Protostars and Planets IV, 457 
\bibitem[Najita et al.(2007)]{na07} Najita, J., Carr, J.~S., Glassgold, A.~E.\& Valenti, J.~A.\ 2007, Protostars and Planets V, 507
\bibitem[Ojha et al.(2011)]{oj11} Ojha, D.~K., Samal, M.~R., Pandey, A.~K., et al.\ 2011, \apj, 738, 156. doi:10.1088/0004-637X/738/2/156
\bibitem[Pomohaci et al.(2017)]{po17} Pomohaci, R., Oudmaijer, R.~D., Lumsden, S.~L., Hoare, M.~G., \& Mendigut{\'{\i}}a, I.\ 2017, \mnras, 472, 3624 
\bibitem[Puga et al.(2010)]{pu10} Puga, E., Mar{\'{\i}}n-Franch, A., Najarro, F., et al.\ 2010, \aap, 517, A2 
\bibitem[Rayner et al.(2016)]{ra16} Rayner, J., Tokunaga, A., Jaffe, D., et al.\ 2016, \procspie, 9908, 990884. doi:10.1117/12.2232064
\bibitem[Rodr{\'\i}guez-Esnard et al.(2014)]{ro14} Rodr{\'\i}guez-Esnard, T., Migenes, V., \& Trinidad, M.~A.\ 2014, \apj, 788, 176
\bibitem[S{\'a}nchez-Monge et al.(2008)]{sa08} S{\'a}nchez-Monge, {\'A}., Palau, A., Estalella, R., et al.\ 2008, \aap, 485, 497. doi:10.1051/0004-6361:20078406
\bibitem[Schreyer et al.(2002)]{sc02} Schreyer, K., Henning, T., van der Tak, F.~F.~S., et al.\ 2002, \aap, 394, 561. doi:10.1051/0004-6361:20021160
\bibitem[Schreyer et al.(2006)]{sc06} Schreyer, K., Semenov, D., Henning, T., et al.\ 2006, \apjl, 637, L129
\bibitem[Science Software Branch at STScI(2012)]{pyraf}Science Software Branch at STScI 2012, PyRAF: Python alternative for IRAF, Astrophysics Source Code Library, ascl:1207.011
\bibitem[Smith et al.(2009)]{sm09} Smith, R.~L., Pontoppidan, K.~M., Young, E.~D., et al.\ 2009, \apj, 701, 163
\bibitem[Snell \& Bally(1986)]{sn86} Snell, R.~L. \& Bally, J.\ 1986, \apj, 303, 683. doi:10.1086/164117
\bibitem[Sridharan et al.(2002)]{sr02} Sridharan, T.~K., Beuther, H., Schilke, P., et al.\ 2002, \apj, 566, 931. doi:10.1086/338332
\bibitem[\protect\astroncite{{Stahler} and {Palla}}{2005}]{Stahler05}
{Stahler}, S.~W. and {Palla}, F.: 2005,
\newblock {\em {The Formation of Stars}},
\newblock Wiley-VCH, Boschstra$\beta$e 12, Weinheim 69469, Germany
\bibitem[Takami et al.(2012)]{ta12} Takami, M., Chen, H.-H., Karr, J.~L., 2012, \apj, 748,8
\bibitem[Tambovtseva et al.(2014)]{ta14} Tambovtseva, L.~V., Grinin, V.~P., \& Weigelt, G.\ 2014, \aap, 562, A104
\bibitem[Tambovtseva et al.(2016)]{ta16} Tambovtseva, L.~V., Grinin, V.~P., \& Weigelt, G.\ 2016, \aap, 590, A97 
\bibitem[Urquhart et al.(2008)]{ur08} Urquhart J. S. et al., 2008a, in Beuther H., Linz H., Henning T., eds, ASP Conf. Ser. Vol. 387, Massive Star Formation: Observations Confront Theory. Astron. Soc. Pac., San Francisco, p. 381
\bibitem[van der Tak et al.(2005)]{va05} van der Tak, F.~F.~S., Tuthill, P.~G., \& Danchi, W.~C.\ 2005, \aap, 431, 993. doi:10.1051/0004-6361:20041595
\bibitem[van der Tak et al.(2006)]{de06} van der Tak, F.~F.~S., Walmsley, C.~M., Herpin, F., et al.\ 2006, \aap, 447, 1011
\bibitem[Van Der Walt et al. (2011)]{numpy}Van Der Walt, S., Colbert, S. C., \& Varoquaux, G. 2011, ArXiv e-prints, arXiv:1102.1523 [cs.MS]
\bibitem[Vacca et al.(2003)]{va03} Vacca, W.~D., Cushing, M.~C., \& Rayner, J.~T.\ 2003, \pasp, 115, 389. doi:10.1086/346193
\bibitem[van Dishoeck et al.(2009)]{di09}van Dishoeck, E. F., van Kempen, T. A., \& Güsten, R. 2009, in ASP Conf. Ser. 417, Submillimeter Astrophysics and Technology: a Symp. Honoring Thomas G. Phillips, ed. D. C. Lis, J. E. Vaillancourt, P. F. Goldsmith et al. (San Francisco, CA: ASP), 203
\bibitem[van Kempen et al.(2009)]{ke09}van Kempen, T. A., van Dishoeck, E. F., Güsten, R., et al. 2009b, \aap, 507, 1425
\bibitem[Virtanen et al. (2020)]{scipy}Virtanen, P., Gommers, R., Oliphant, et al. (2020). SciPy 1.0: Fundamental Algorithms for Scientific Computing in Python. Nature Methods.
\bibitem[Wang et al.(2012)]{wa12} Wang, K.-S., van der Tak, F.~F.~S., \& Hogerheijde, M.~R.\ 2012, \aap, 543, A22. doi:10.1051/0004-6361/201117044
\bibitem[Wang et al.(2013)]{wa13} Wang, K.-S., Bourke, T.~L., Hogerheijde, M.~R., et al.\ 2013, \aap, 558, A69
\bibitem[Wallace \& Hinkle(1997)]{wa97} Wallace, L., \& Hinkle, K.\ 1997, \apjs, 111, 445
\bibitem[Wolfire \& Cassinelli(1984)]{wo84} Wolfire, M.~G., \& Cassinelli, J.~P.\ 1984, \baas, 16, 960 
\bibitem[Xu et al.(2012)]{xu12} Xu, J.-L., Wang, J.-J., \& Qin, S.-L.\ 2012, \aap, 540, L13
\bibitem[Yang et al.(2010)]{ya10} Yang, B., Stancil, P.~C., Balakrishnan, N., et al.\ 2010, \apj, 718, 1062
\bibitem[Y{\i}ld{\i}z et al.(2013)]{yi13} Y{\i}ld{\i}z, U.~A., Kristensen, L.~E., van Dishoeck, E.~F., et al.\ 2013, \aap, 556, A89
\bibitem[Zinchenko et al.(2012)]{zi12} Zinchenko, I., Liu, S.-Y., Su, Y.-N., et al.\ 2012, \apj, 755, 177
\bibitem[Zinchenko et al.(2015)]{zi15} Zinchenko, I., Liu, S.-Y., Su, Y.-N., et al.\ 2015, \apj, 810, 10
\end{thebibliography}

\software{Science Software Branch at STScI 2012, PyRAF: (\citealt{pyraf}), Numpy (\citealt{numpy}), Scipy (\citealt{scipy}), Matplotlib (\citealt{matplotlib})}
\end{document}